\documentclass[submission, Phys]{SciPost}

\hypersetup{
	colorlinks,
	linkcolor={red!50!black},
	citecolor={blue!50!black},
	urlcolor={blue!80!black}
}

\usepackage[english]{babel}

\usepackage[utf8]{inputenc} 
\usepackage{lmodern} 
\usepackage[T1]{fontenc}
\usepackage{arydshln}
\usepackage{amssymb}
\usepackage{mathtools} 
    \numberwithin{equation}{section}

\usepackage{bbm} 

\usepackage{verbatim} 

\usepackage{graphicx}
\usepackage{subfig}

\usepackage{cleveref} 

\usepackage{tikz}
	\usetikzlibrary{decorations.pathreplacing}
	\usetikzlibrary{decorations.pathmorphing} 
	\usetikzlibrary{matrix,arrows} 
    \usetikzlibrary{fadings} 
\usetikzlibrary{arrows.meta}

\definecolor{cellgreen}{RGB}{132,224,0}

\definecolor{cellblue}{RGB}{124,202,239}

\definecolor{nodeblue}{RGB}{91,139,232}

\usepackage{multirow} 

\usepackage{enumitem}

\usepackage{scalerel,stackengine} 
    \newcommand\pig[1]{\scalerel*[5.5pt]{\Big#1}{%
    \ensurestackMath{\addstackgap[1.5pt]{\big#1}}}}
    \newcommand\pigl[1]{\mathopen{\pig{#1}}}
    \newcommand\pigr[1]{\mathclose{\pig{#1}}}

\DeclarePairedDelimiter{\bra}{\langle}{\rvert}
\DeclarePairedDelimiter{\ket}{\lvert}{\rangle}
\DeclarePairedDelimiterX{\ketbra}[2]{\lvert}{\rvert}{#1\rangle \langle#2}
\DeclarePairedDelimiterX{\braket}[2]{\langle}{\rangle}{#1\vert#2}

\DeclarePairedDelimiter{\cket}{\lvert}{\rangle\!\rangle}
\DeclarePairedDelimiterX{\cbraket}[2]{\langle\!\langle}{\rangle}{#1\vert#2}
\DeclarePairedDelimiterX{\bracket}[2]{\langle}{\rangle\!\rangle}{#1\vert#2}
\DeclarePairedDelimiterX{\cbracket}[2]{\langle\!\langle}{\rangle\!\rangle}{#1\vert#2}

\usepackage{bm} 
\newcommand{\vect}[1]{\bm{{#1}}}
	
\newcommand{\I}{\mathrm{i}}
\newcommand{\E}{\mathrm{e}}

\DeclareMathOperator*{\ordprod}{\prod\limits^{\vbox to -.5ex{\kern-0.5ex\hbox{$\leftharpoonup$}\vss}}}
\DeclareMathOperator*{\ordprodopp}{\prod\limits^{\vbox to -.5ex{\kern-0.5ex\hbox{$\rightharpoonup$}\vss}}}

\newcommand{\rmi}{\mathrm{i}}

\begin{document}

\begin{center} {\Large \textbf{
    Norms, overlaps and Yangian descendants \\[.2ex] for the Haldane--Shastry spin chain
}
} 
\end{center}

\begin{center}
    Yunfeng Jiang\textsuperscript{1,2},
    Jules Lamers\textsuperscript{3},
    Yuan Miao\textsuperscript{4\,$\curvearrowright$\,5\,$\diamond$}
\end{center}

\begin{center}
    \textsuperscript{\textbf{1}} School of Physics \& Shing-Tung Yau Center, Southeast University,\\ 
    Nanjing  211189, P. R. China \\
    \textsuperscript{\textbf{2}} Peng Huanwu Center for Fundamental Theory, Hefei, Anhui 230026, P. R. China \\[1ex]
    \textsuperscript{\textbf{3}} School of Mathematics and Statistics, University of Glasgow \\ University Place, Glasgow G12 8QQ, UK \\[1ex]
    \textsuperscript{\textbf{4}} Kavli Institute for the Physics and Mathematics of the Universe (WPI), \\ UTIAS, The University of Tokyo, Kashiwa, Chiba 277-8583, Japan \\
    \textsuperscript{\textbf{5}} Laboratoire de Physique de l'École Normale Supérieure, ENS, Université PSL, CNRS, Sorbonne Université, Université Paris Cité, Paris, France \\
    \textsuperscript{$\diamond$} \href{mailto:yuan.miao.physics@gmail.com} {\texttt{yuan.miao.physics@gmail.com}} 
    \bigskip
\end{center}

\begin{center}
	\today
\end{center}

\section*{Abstract}

The Haldane--Shastry spin chain is a prototypical integrable model with long-range interactions, notable for hosting quasiparticles with fractional statistics and serving as a discrete analogue of a conformal field theory. Its remarkable simplicity is closely tied to a full Yangian spin symmetry. While the highest-weight states for this symmetry are known explicitly, a systematic treatment of the descendant states, needed for the computation of various physical quantities, has remained incomplete. 
In this work, we provide a detailed construction of these descendants in terms of the algebraic Bethe ansatz following recent work of Ferrando {et al}. In the limit of extreme twist, it includes the Gelfand--Tsetlin basis. As an application, we derive product and determinant formulae for norms and overlaps of these states.

\vspace{10pt}
\noindent\rule{\textwidth}{1pt}
\tableofcontents\thispagestyle{fancy}
\noindent\rule{\textwidth}{1pt}
\vspace{10pt}

\section{Introduction}
\label{sec:intro}

The Haldane--Shastry~(HS) spin chain~\cite{Haldane_1988, Shastry_1988} is the prototypical example of an integrable long-range interacting spin chain. It possesses many remarkable properties, such as hosting quasiparticles with fractional statistics. It has intimate connections with 2-dimensional conformal field theories~(\textsc{cft}), and in particular can be viewed as a discrete version of the level-1 $SU(2)$ Wess--Zumino--Witten~(WZW) model~\cite{HHTBP,Bernard:1994wg, Schoutens:1994au,Bouwknegt:1994sj,Bouwknegt:1994bk}. 

While one would generally expect a model with long-range interactions to be more complicated than spin chains with nearest-neighbour interactions, the HS chain is in fact in some respects simpler than the Heisenberg \textsc{xxx} chain. Indeed, the HS chain is not just \emph{exactly} but even \emph{explicitly} solvable: its eigenvalues and (certain) eigenvectors can be written in a simple, closed form \cite{Haldane_1991_spinon_gas, HHTBP, BGHP}. This simplicity also shows up in an enlarged spin symmetry: the Hamiltonian is invariant under the \emph{full} Yangian \cite{HHTBP, BGHP}, which contains the standard $\mathfrak{su}_2$ spin-symmetry algebra as a subalgebra. The monodromy matrix for the HS chain is different from the one of the Heisenberg \textsc{xxx} chain~\cite{HHTBP,BGHP}, and the Hamiltonian comes from expansions of the so-called quantum determinant. Thanks to this Yangian symmetry, the Hilbert space of the HS chain decomposes into several Yangian multiplets (irreducible representations) that are the eigenspaces of the HS Hamiltonian. As a consequence, the HS chain has high degeneracies, as was already observed in \cite{Haldane_1988}. The Yangian highest-weight (\textsc{yhw}) states, also known as pseudovacua, of these multiplets have been studied extensively \cite{HHTBP,BGHP}, and their wave functions are given explicitly in terms of Jack polynomials (see Sec.~\ref{sec:overview} for details). Together, the explicit \textsc{yhw} states and the Yangian symmetry in principle determine all eigenstates. However, the monodromy matrix of the HS chain is complicated, requiring one to work with the more general setting of spin-Calogero--Sutherland systems, from which the HS chain is obtained by a process called freezing~\cite{Polychronakos:1992ki, BGHP, Talstra_1995, Liashyk:2024ekz}. Perhaps for this reason, the remaining (Yangian descendant) eigenstates have received less attention, having only been mentioned briefly in various places~\cite{HHTBP, BGHP, TU, FLLS}. 

The ground-state correlation functions of the HS chain~\cite{Vollhardt_1988, Haldane_Zirnbauer_1993, Yamamoto:1999gh, Kuramoto_2009, Enciso_2012, Stephan_2017, Nielsen:2011py, Bulchandani_2024} and the single particle Green's functions and dynamical correlation functions of the CS system~\cite{Lesage:1994ii, Zirnbauer:1995zz, Serban:1995zp, Ha:1994xj, Ha:1994nep} and its spin-generalisation~\cite{Uglov:1997ia} have received much attention. In comparison, the finite-temperature correlations of the HS chain have remained under-explored, but see~\cite{Peysson_2003}. 
In order to study out-of-equilibrium dynamics and finite-temperature properties of the HS chain, which require sums over all states in the Hilbert space, a better understanding of the Yangian descendants is required.

\textbf{The goal of this paper} is to determine the norms and overlaps of the \textsc{yhw} and descendant states, paving the way for the study of various important physical properties, including the correlation functions and out-of-equilibrium dynamics.

Since the monodromy matrix of the HS chain~\cite{HHTBP,BGHP} \emph{commutes} with the Hamiltonian, one cannot use conventional Bethe-ansatz techniques in the same way as for the Heisenberg \textsc{xxx} chain. Instead, like any finite-dimensional Yangian multiplet, each HS eigenspace has the structure of an `effective' inhomogeneous Heisenberg chain with inhomogeneities fixed at special values that depend on the eigenspace. Building on \cite{TU}, in \cite{FLLS} it was proposed to study the complete eigenbasis using an `internal Bethe ansatz' inside each eigenspace separately, by diagonalizing the HS Hamiltonian simultaneously with the transfer matrix. This allows one to leverage the powerful toolkit of traditional Heisenberg-style integrability \emph{per eigenspace}. Notably, the joint eigenstates are given by the algebraic Bethe ansatz (\textsc{aba}), leading to Bethe-ansatz equations (\textsc{bae}) for each eigenspace, or, equivalently, $TQ$-relations. When the parameters in the \textsc{aba} are `on shell', i.e.\ solve the \textsc{bae}, the Bethe states become eigenstates of the transfer matrix. Considering the (diagonally) twisted transfer matrix $\mathbf{T}(x; \kappa) = \kappa \, \mathbf{A}(x) + \kappa^{-1} \, \mathbf{D}(x)$, this yields a description of the descendants as an eigenbasis depending on the value of the twist $\kappa$, orthogonal in the regime of $\kappa$ where the transfer matrix is hermitian. Among all possible twists, two choices are special: 
\begin{itemize}
    \item in the periodic case $\kappa=1$, the effective spin chain has $
    \mathrm{SU}(2)$ symmetry;
    \item in the limits $\kappa\to 0$ or $\kappa\to\infty$ of extreme twist, the Bethe roots become explicit, and the Bethe vectors reduce to the so-called Gelfand--Tsetlin basis~\cite{GelfandTsetlin1950, Nazarov_1994, Molev_1994, molev2007yangians}.
\end{itemize}
The \textsc{aba} allows one to compute the overlaps of descendant states, which are orthogonal when the transfer matrix is hermitian and have norms given by the Gaudin determinant~\cite{Gaudin_2014, Gaudin_1981, Korepin:1982gg, Kostov:2012yq}. Furthermore, overlaps of one `on-shell' and one `off-shell' Bethe state are given by the Slavnov determinant formula~\cite{Slavnov:1989uvz, Kostov:2012yq}. 

There is an alternative approach that uses \textsc{cft} techniques, by expressing the HS wave functions in terms of correlators of $SU(2)_1$ WZW \textsc{cft}~\cite{Cirac:2010rka, Herwerth:2015pga}, in order to construct the complete eigenstates of the HS chain. From this construction, orthogonality and norms are not obvious. 

\medskip

\textbf{Our main results} are as follows. The `quantum numbers' labelling the eigenspaces of the HS chain are called `motifs', see Sec.~\ref{sec:motifs}. Their \textsc{yhw} states are known explicitly, see Sec.~\ref{subsec:YHW}. In Sec.~\ref{subsec:descendantnorm}, we use the orthogonality of Jack polynomials to prove that the \textsc{yhw} states for different motifs $\mu$ are orthogonal (despite `accidental' degeneracies of the HS chain), and that the \textsc{yhw} state $\ket{0}_{\mu}$ for the motif $\mu$ has norm-squared given by the simple factorised formula 
\begin{equation} \label{eq:normHigh}
    {_{\mu}}\braket{0}{0}_{\mu} = N^M \prod_{m < m'}^M \frac{\mu_m -\mu_{m'}-1}{\mu_m - \mu_{m'} + 1} \; ,
\end{equation}
where $M$ is the number of down spins in $\ket{0}_{\mu}$. In Sec.~\ref{subsec:textsc{aba}} we obtain Yangian descendant states by diagonalising the twisted transfer matrix in the Yangian multiplet labelled by $\mu$ using the \textsc{aba} $\ket{\vect{u}}_{\mu} \coloneqq \mathbf{B}(u_1)\cdots \mathbf{B}(u_K)\,\ket{0}_\mu$, leading to the \textsc{bae}~\eqref{eq:effectiveBAE}. In Sec.~\ref{subsec:descendantnorm} we show that the norm of Yangian descendant states is proportional to the Gaudin determinant $G_{\vect{u}, \vect{u}}$ given in~\eqref{eq:Gaudin},
\begin{equation}
    \frac{{_\mu}\braket{\vect{u}}{\vect{u}}_{\mu}}{{_\mu}\braket{0}{0}_{\mu} } 
    = \biggl( \, \prod_{k=1}^K A_\mu (u_k) \, D_\mu (u_k) \biggr) \, G_{\vect{u}, \vect{u}} \; ,
\end{equation}
while the overlap between on-shell and off-shell Yangian descendants (within the same Yangian multiplet) is proportional to the Slavnov determinant $S_{\vect{v}, \vect{u}}$ from \eqref{eq:Slavnov},
\begin{equation}
    \frac{{_\mu}\braket{\vect{v}}{\vect{u}}_{\mu}}{{_\mu}\braket{0}{0}_{\mu}}  = \delta_{K,K'} \Biggl( \, \prod_{k=1}^K A_\mu (v_k) \, D_\mu (u_k) \Biggr) \, S_{\vect{v}, \vect{u}} \; ,
\end{equation}
with proportionality factors containing polynomials $A_\mu (x)$ and $D_\mu (x)$ that are given in \eqref{eq:Aeigenvaluehw}--\eqref{eq:Deigenvaluehw}.
\medskip

\textbf{The rest of the paper} is structured as follows. In Sec.~\ref{sec:overview}, we review various known properties of the HS chain and present our main results in more detail (see the end of Sec.~\ref{subsec:YHW}, and Sec.~\ref{subsec:Ydescendant}). The proofs of these results are given in the subsequent two sections. In Sec.~\ref{sec:CSbasis}, we review the relationship between the HS chain and the spin Calogero--Sutherland system through `freezing', and explain how to formulate the \textsc{aba} for the HS chain using `frozen' Dunkl operators. Equipped with this formalism, we show in Sec.~\ref{sec:HSABA} how to construct Yangian descendant states via the \textsc{aba} technique. Then we discuss the special values of the twist, and derive the norm and overlap formulae. To illustrate our results, we present explicit examples in Sec.~\ref{sec:example}. We conclude and discuss future directions in Sec.~\ref{sec:conclusion}. In addition, there are two appendices: App.~\ref{app:ABArelations} contains the Yangian relations for the quantum operators for easy reference, and App.~\ref{app:inhomogeneousXXX} contains the basics of the inhomogeneous Heisenberg chain and TQ relations.

\section{Overview: old, and some new, results for Haldane--Shastry}
\label{sec:overview}

We start with the Hilbert space $(\mathbb{C}^2)^{\otimes N}$ for $N$ sites with spin $1/2$, and use the standard notation for local operators, \emph{e.g.}\ $\mathbf{O}_j = \mathbb{I}^{\otimes (j-1)} \otimes \mathbf{O} \otimes \mathbb{I}^{\otimes (N-j)}$. In particular, $\sigma^\alpha_j$ denotes the $\alpha$th Pauli matrix acting locally at the $j$th site. The permutation operator can be defined as
\begin{equation} \label{eq:perm}
    \mathbf{P}_{ij} = \frac{1}{2} \left( 1 + \vec{\sigma}_i \cdot \vec{\sigma}_j \right) \, , 
\end{equation}
such that $\mathbf{P}_{ij} \, \mathbf{O}_i = \mathbf{O}_j \, \mathbf{P}_{ij}$ and $\mathbf{P}_{ij}^2 = 1$.
The Haldane--Shastry spin chain is defined by the Hamiltonian
\begin{equation} \label{eq:H_HS}
    \mathbf{H}_\textsc{hs} = \sum_{i < j}^N \frac{1 - \mathbf{P}_{ij}}{4 \sin^2\bigl(\pi (i - j)/N\bigr)} \; .
\end{equation}

\subsection{Conserved charges} 
\label{sec:conserved_charges}

The HS chain has two types of conserved charges. There are $N$ `basic' charges, containing the translation operator (or momentum) and Hamiltonian \cite{Inozemtsev_1990, HHTBP, Talstra_1995}. These commute with all Yangian generators; from the Yangian perspective, these charges are generated by the quantum determinant \cite{BGHP}. Higher conserved charges can be derived systematically \cite{Talstra_1995}. For instance,
\begin{equation}
\begin{aligned}
    \mathbf{Q}_3 = &\,\sum_{\substack{i,j,k=1\\ i\neq j\neq k}}^{N} \frac{\mathbf{P}_{ij} \, \mathbf{P}_{jk}}{\sin\bigl[\frac{\pi}{N}(i-j)\bigr] \, 
    \sin\bigl[\frac{\pi}{N}(j-k)\bigr] \, \sin\bigl[\frac{\pi}{N}(k-i)\bigr]} \; ;\\
    \mathbf{Q}_4 = &\,\sum_{\substack{i,j,k,l=1\\ i\neq j\neq k\neq l}}^N
    \frac{\mathbf{P}_{ij} \, \mathbf{P}_{jk} \, \mathbf{P}_{kl}}{\sin\bigl[\frac{\pi}{N}(i-j)\bigr] \, \sin\bigl[\frac{\pi}{N}(j-k)\bigr] \, \sin\bigl[\frac{\pi}{N}(k-l)\bigr] \, \sin\bigl[\frac{\pi}{N}(l-i)\bigr]}\\
    & \quad - 2\sum_{\substack{i,j=1\\ i\ne j}}^N \frac{\mathbf{P}_{ij}}{\sin^4\bigl[\frac{\pi}{N}(i-j)\bigr]} \; .
\end{aligned}
\end{equation}
The literature on the HS chain has traditionally focussed on these commuting charges possessing Yangian symmetry. 

On top of this, one can construct additional charges \cite{Minahan_1993, FLLS}. In \cite{FLLS}, `Heisenberg-style' charges were constructed that depend on a twist parameter, and commute with each other and the `basic' charges, but not with all Yangian generators. They arise from a twisted transfer matrix, 
$\mathbf{T}(x; \kappa) = \kappa \, \mathbf{A}(x) + \kappa^{-1} \, \mathbf{D}(x)$,
in analogy to the Heisenberg \textsc{xxx} chain. An explicit example of such a Heisenberg-style charge for arbitrary twist $\kappa$ is
\begin{equation} \label{eq:Heis_style}
    \frac{\kappa + \kappa^{-1}}{2} \, \sum_{i<j}^N \mathbf{P}_{ij} + \frac{\kappa - \kappa^{-1}}{4\,\I} \, \sum_{i<j}^N \frac{\E^{\I\pi(i-j)/N} \, \sigma^z_j - \E^{\I\pi(j-i)/N} \, \sigma^z_i}{\sin\bigl[\frac{\pi}{N}(i-j)\bigr]} \, \mathbf{P}_{ij} \, .
\end{equation}
At $\kappa=1$ this reduces to the total spin operator (quadratic Casimir). In that case, a nontrivial example is
\begin{equation} \label{eq:Heis_style_isotr}
    \sum_{\substack{i,j,k=1\\ i\neq j\neq k}}^{N}  \biggl( \frac{1}{3} + \I\,\cot\Bigl[\frac{\pi}{N}(i-j)\Bigr]\biggr) \, \mathbf{P}_{ij} \, \mathbf{P}_{jk} \, .
\end{equation}
The presence of these additional charges is related to superintegrability. For more, see \cite{FLLS}.

\subsection{Motif description of the spectrum: eigenvalues} 
\label{sec:motifs}

The spectrum of the HS chain can be described conveniently by simple combinatorial objects called motifs. A \emph{motif} is a sequence $\mu =(\mu_1,\ldots,\mu_M)$  of integers satisfying the exclusion rule \cite{Haldane_1988}
\begin{equation} \label{eq:motifs}
    \mu_{m+1} > \mu_m + 1 \, , \qquad 1\leqslant \mu_m\leqslant N-1 \, ,
\end{equation}
where $N$ is the length of the spin chain. Motifs label eigenspaces.

When we consider the complete eigenbasis of the HS chain, it will be useful to have a notation for the set of all motifs. We write $\mathcal{M}_N$ for the set of all motifs for a chain of length~$N$. For example: $\mathcal{M}_2=\{\varnothing, (1 ) \}$, $\mathcal{M}_3=\{\varnothing, (1), (2 ) \}$, $\mathcal{M}_4=\{\varnothing, (1), (2), (3), (1,3) \}$. Note that $\mathcal{M}_{N-1} \subset \mathcal{M}_N$. The motifs in $\mathcal{M}_N \setminus \mathcal{M}_{N-1}$ are precisely those obtained by appending the integer $N-1$ to every motif in $\mathcal{M}_{N-2}$. This recursion implies that the number of motifs follows a Fibonacci sequence, $\{\#\mathcal{M}_N\}_{N\geqslant 2} = \{2, 3, 5, 8, 13, \ldots\}$.

\paragraph{Energy and momentum.} 
Consider the rescaled motif
\begin{equation} \label{eq:p_m}
    p_m = \frac{2\pi}{N} \, \mu_m \, , \qquad m = 1,\dots,M\,.
\end{equation}
Then the eigenspace associated with the motif $\mu = (\mu_1, \dots, \mu_M)\in\mathcal{M}_N$ has (total) momentum $p(\mu)$ and energy $E(\mu)$ given in terms of the motif as
\begin{equation} \label{eq:EP}
    p(\mu)= \sum_{m=1}^M p_m  \ \, \mathrm{mod}\ 2\pi \; , \quad
    E(\mu)= \sum_{m=1}^M \varepsilon(p_m)\; ,
\end{equation}
where the dispersion relation is
\begin{equation}
\label{eq:disperseRel}
    \varepsilon(p_m) = \frac{N^2}{8\pi^2} \ p_m \, (2\pi - p_m) \;.
\end{equation}

Thanks to \eqref{eq:EP}, the $p_m$ from \eqref{eq:p_m} can be interpreted as on-shell quasimomenta of the quasiparticles (magnons). According to \eqref{eq:p_m}, all quasimomenta are real for \textsc{yhw} states. Their values are exactly quantised as for free particles on a circle, with the motif rule~\eqref{eq:motifs} a generalised Pauli principle enforcing a fractional (exclusion) statistics~\cite{Haldane_1991_spinon_gas}.

Since the dispersion \eqref{eq:disperseRel} is quadratic and the energies \eqref{eq:EP} are additive even \emph{on shell}, the spectrum of the HS chain is extremely simple: the energies are fully explicit and, up to an overall normalisation, take integer values. This stands in contrast to typical Bethe-ansatz solvable models, such as the Heisenberg chain, where solving the \textsc{bae} is non-trivial and the spectrum generally lacks a closed-form expression.

\paragraph{Comparison to BAE.}
Consider for a moment quasimomenta $p_m$ satisfying the following Bethe-ansatz-like equations~\cite{Haldane_1991_spinon_gas, Ha1993} 
\begin{equation}
\label{eq:simpleBAE}
    N \,p_m = 2\pi \, I_m + 2\pi \!\! \sum_{\substack{n=1 \\ n \neq m}}^M \! \mathrm{sgn}(p_m - p_n) \; ,
\end{equation}
where the quantum numbers $I_m$ take ascending values in $\{\frac{M+1}{2}, \frac{M+3}{2}, \dots, N-\frac{M+1}{2}\}$. 

From \eqref{eq:simpleBAE}, the scattering phase between quasiparticles is nearly trivial, $S(p_m, p_n) = e^{\mathrm{i}\,\pi\,\text{sgn}(p_m - p_n)} = -1$, indicating a `free spectrum'\footnote{\ The dressed scattering phase for spinons, i.e.\ the excitations over the antiferromagnetic ground state, is also momentum-independent~\cite{Essler_1995}, indicating the free semionic gas nature of the HS chain~\cite{Haldane_1994}.}. The solutions to \eqref{eq:simpleBAE} are
\begin{equation}
    p_m = \frac{2\pi}{N} \left( I_m +m - \frac{M+1}{2} \right) \; ,
\end{equation}
with quantum numbers $I_m$ in an ascending order. Defining $\mu_m = I_m +m - (M+1)/2$ we are led to integers increasing according to the exclusion rule \eqref{eq:motifs}. Thus, the motifs can be viewed as an analogue of Bethe quantum numbers in the Heisenberg chain.

We stress that \eqref{eq:simpleBAE} are not Bethe-ansatz equations in the usual sense, in that they do not follow from the periodicity of any Bethe-ansatz type wave function. However, \eqref{eq:simpleBAE} do arise as the long-range limit of the asymptotic Bethe ansatz equations of the Inozemtsev spin chain as $N\to \infty$, cf. \textsection4.4 of \cite{KL_Ino}.

\subsection{Yangian symmetry} 
\label{subsec:yangian}

Each motif is in general associated with more than one eigenstate, \textit{i.e.}\ the spectrum is degenerate. This degeneracy arises from the underlying symmetry of the model. For the HS chain, the degeneracies are much larger \cite{Haldane_1988} than one would expect based on the form of the Hamiltonian \eqref{eq:H_HS}. 
 
Clearly, the HS chain is isotropic: the Hamiltonian commutes with the $\mathfrak{su}_2$ generators
\begin{align} \label{eq:Salpha}
    \mathbf{S}^{\alpha}=\frac{1}{2}\sum_{j=1}^N\sigma_j^{\alpha} \, ,\qquad \alpha\in \{ x,y,z\} \; .
\end{align}
In fact, the model has much more spin symmetry, which turns out to be governed by the following algebraic object.

\paragraph{Yangian symmetry.} 
The Yangian is an infinite-dimensional quantum group generated by the \eqref{eq:Salpha} together with a second set of `affine generators'. For the HS chain, they are given by \cite{HHTBP} (see also \cite{Bernard:1994wg,LS})
\begin{align} \label{eq:Qalpha}
    \mathbf{Q}^{\alpha} = \sum_{i<j}^N \cot\biggl(\frac{\pi}{N}(i-j)\biggr) \left(\vec{\sigma}_i\times\vec{\sigma}_j\right)^{\alpha} \, ,\qquad \alpha\in \{ x,y,z\} \; .
\end{align}
These bilocal operators form an adjoint representation of $\mathfrak{su}_2$,
\begin{align}
    \left[ \mathbf{S}^{\alpha}, \mathbf{Q}^{\beta} \right]=\sum_{\gamma \in \{ x,y,z\}} 2 \, \mathrm{i} \, \epsilon^{\alpha\beta\gamma} \, \mathbf{Q}^{\gamma} \; ,
\end{align}
and obey additional Serre-type relations~\cite{Drinfeld:1986in, molev2007yangians}. Higher Yangian generators are obtained from repeated commutators of $\mathbf{Q}^{\alpha}$.

Like the spin operators \eqref{eq:Salpha}, the generators \eqref{eq:Qalpha} are symmetries of the HS chain \cite{HHTBP,BGHP}: 
\begin{equation}
    [\mathbf{S}^{\alpha},\mathbf{H}_\textsc{hs}] = [\mathbf{Q}^{\alpha},\mathbf{H}_\textsc{hs}] = 0 \, .
\end{equation}
Due to this large amount of symmetry, the Hilbert space decomposes into eigenspaces that form Yangian multiplets, \textit{i.e.}\ irreducible highest-weight representations of the Yangian. Each such eigenspace is labelled by a motif.

Other sets of generators for the Yangian are available as well. This in particular includes the monodromy matrix\cite{BGHP}, whose construction for the HS chain we will review in Sec.~\ref{sec:CSbasis}. This gives access to the framework of the Quantum Inverse Scattering Method~(QISM), which we will exploit in Sec.~\ref{sec:HSABA}.

\paragraph{Degeneracies.} 
Finite-dimensional $\mathfrak{su}_2$-multiplets are characterised by their (total) spin $s\,(s+1)$, or equivalently by the (half-integer) spin $s$, for a multiplet of dimension $2\,s+1$. Similarly, any finite-dimensional Yangian multiplet is characterized by a quantity called the `Drinfeld polynomial' (cf.\ \cite{molev2007yangians}), which is a polynomial in the spectral parameter. For the multiplet labelled by a motif $\mu \in \mathcal{M}_N$, the Drinfeld polynomial is \cite{BGHP,LPS}
\begin{equation}
    P_\mu(x) = \! \prod_{\substack{n = 1 \\ n \notin  \mu \cup (\mu +1)}}^N \!\!\!\!\!\!\!\! \biggl(x + \frac{N+1-2\,n}{2} \biggr) \; ,
    \label{eq:Drinfeldpoly}
\end{equation}
where we introduced the `thickened motif' $\mu \cup (\mu +1) = ( \mu_1,\mu_1 + 1,\dots,\mu_M , \mu_M +1 )$.

In particular, the Drinfeld polynomial determines the dimension of the multiplet. This works as follows, see e.g.~\cite{Chari_1991}. The zeroes of any Drinfeld polynomial can be organized into `strings'.\footnote{\ Note that this notion of `$\ell$-string' differs from the concept with the same name in the context of complex Bethe roots and bound states in the Heisenberg chain.} By definition, an `$\ell$-string' is a maximal set of consecutive zeroes $\{ z , z+1 , \dots, z+\ell-1 \}$ of the Drinfeld polynomial. For \eqref{eq:Drinfeldpoly} associated to the motif $\mu$, defining $\mu_0=-1$ and $\mu_{M+1} = N+1$, the strings of zeroes $\{ \mu_{m} +2 -\tfrac{N+1}{2} , \dots , \mu_{m+1} -1 -\tfrac{N+1}{2}\}$ (possibly empty) with
\begin{equation} \label{eq:string_lengths}
    \ell_m = \mu_{m+1} - \mu_m -2 \; , \qquad m = 0 , 1 , \dots , M\; ,
\end{equation}
In terms of usual spin multiplets, each $\ell_m$-string corresponds to a spin-$\frac{\ell_m}{2}$ irrep of $\mathfrak{sl}_2$. In particular, a 0-string corresponds to a singlet (trivial representation). The full Drinfeld polynomial corresponds to the tensor-product representation whose factors have spins $\frac{\ell_0}{2}, \frac{\ell_1}{2}, \dots, \frac{\ell_M}{2}$ in terms of $\mathfrak{sl}_2$. Thus, by \eqref{eq:Drinfeldpoly} the degeneracy of the multiplet labelled by the motif $\mu$ is
\begin{equation} \label{eq:dimensionmu}
    \begin{aligned}
    \text{dim}(\mu) = \prod_{m=0}^{M} \! ( \ell_m +1 ) = \left\{\begin{array}{ll} N+1 &\text{if }\mu=\varnothing \; ,\\ \displaystyle \mu_1 \, (N-\mu_M)\prod_{m=1}^{M-1}(\mu_{m+1}-\mu_m-1) &\text{if }M\geqslant 1 \; . \end{array}\right.
    \end{aligned}
\end{equation}
While the multiplet is irreducible for the Yangian, as soon there is more than one $\ell$-string, for $\mathfrak{sl}_2$ it decomposes into several spin multiplets, via the usual Clebsch--Gordan decomposition. We will give some examples of this in Sec.~\ref{sec:example}.

In particular, since adding the dimensions~\eqref{eq:dimensionmu} (number of linearly independent Yangian descendants) for all motifs $\mu \in \mathcal{M}_N$ (labelling different eigenspaces) gives the dimension $2^N$ of the whole HS Hilbert space, the preceding description of the energy spectrum is complete.

Each Yangian multiplet consists of a Yangian highest-weight state and its descendants, which we will discuss in turn.

\subsection{Yangian highest-weight states}
\label{subsec:YHW}

Each eigenspace, labelled by some motif $\mu$, contains a unique Yangian highest-weight (\textsc{yhw}) state, or pseudovacuum, $\ket{0}_\mu$.

\paragraph{Explicit form.} 
For a motif $\mu = (\mu_1,\dots,\mu_M)$ the \textsc{yhw} state has magnon number (number of $\downarrow$s) equal to $M$. For instance, for the empty motif $\mu=\varnothing$ the vector $\ket{0}_\varnothing = \ket{{\uparrow}\cdots \uparrow}$ is the ferromagnetic \textsc{yhw} state.

In the coordinate basis, defined by
\begin{align} \label{eq:coord_basis}
    \cket{ n_1, \ldots, n_M } \coloneqq \sigma_{n_1}^- \cdots \sigma_{n_M}^- \ket{\uparrow \cdots \uparrow} \; ,
\end{align}
the $M$-magnon \textsc{yhw} state
\begin{align}
    \ket{0}_{\mu} = \sum_{n_1 < \cdots < n_M}^N \!\!\!\!\! \Psi_{\mu}(n_1, \ldots, n_M) \, \cket{ n_1, \ldots, n_M } 
\label{eq:Y_hw}
\end{align}
labelled by the motif $\mu$ has explicit wave function \cite{Haldane_1991_spinon_gas}
\begin{equation}
    \Psi_{\mu} (n_1, \dots , n_M) = 
    \omega^{n_1} \cdots \omega^{n_M} \, V( \omega^{n_1} , \dots, \omega^{n_M})^2 
    \; \mathsf{P}_{\nu(\mu)} (\omega^{n_1} , \dots, \omega^{n_M}) \; .
    \label{eq:Y_hw_wf}
\end{equation}
The right-hand side is a product of polynomials in $M$ variables that are `evaluated' at roots of unity $\omega^n = \E^{2\pi \rmi \mspace{1mu} n/N}$. The first part of \eqref{eq:Y_hw_wf}, including the Vandermonde determinant
\begin{equation}
    V(z_1, \dots, z_M) = \, \det\limits_{1\leqslant m,m'\leqslant M} \pigl( z_{m}^{M-m'} \pigr)= \prod_{m<m'}^M \!\! ( z_{m} - z_{m'} ) \, ,
\end{equation}
is a Slater--Jastrow wave function multiplied by Marshall signs,
\begin{equation} \label{eq:SlaterJastrow}
    \omega^{n_1} \cdots \omega^{n_M} \, V( \omega^{n_1} , \dots, \omega^{n_M})^2 = (-1)^{M(M-1)/2} \, \E^{\frac{2\pi \I M}{N}\sum_m n_m} \! \prod_{m<m'}^M \!\! 4 \sin^2 \Bigl( \frac{\pi}{N} (n_m - n_{m'}) \Bigr)\; .
\end{equation}
Next, $\mathsf{P}_\nu(z_1, \dots, z_M)$ denotes a Jack polynomial with parameter $\alpha = 1/2$ (also known as a spherical zonal polynomial) and partition $\nu = \nu(\mu)$ determined by the motif through\,%
\footnote{\ Compared to \cite{LPS}, we have used the property $P_{\lambda+1}(z_1,\dots,z_M) = z_1 \cdots z_M \, P_\lambda(z_1,\dots,z_M)$ of Jack polynomials to decrease all parts of the partition by $1$.}
\begin{equation}
    \nu_m = \mu_{M -m+1} - 2 \, (M - m) -1 \; , \quad m \in \{1,2,\cdots , M \} \; .
\label{eq:nu_from_mu}
\end{equation}

For motifs of the form $\mu = (1,3,\dots,2M-1)$, $\nu(\mu)=0$ and $\mathsf{P}_\nu(z_1, \dots, z_M)=1$. In particular, when $N$ is even, the antiferromagnetic ground state at the equator $M=N/2$ (labelled by the `fully packed' motif $\mu = (1,3,\dots,N-1)$) has Slater--Jastrow wave function. This remarkably simple ground state, reminiscent of the fractional quantum Hall effect, is a wave function of the HS chain by design~\cite{Haldane_1988, Shastry_1988}. More about Jack polynomials can be found in e.g.~\cite{Stanley_1989}. 

Other explicit examples and properties of these wavefunctions can be obtained from e.g.\ \textsection 1.1.3 of \cite{LPS}.

\paragraph{Orthogonality and norms.}
The orthogonality of the \textsc{yhw} states is guaranteed by the properties of Jack polynomials: for any two motifs $\mu,\mu' \in \mathcal{M}_N$ with cardinality $M$ and $M'$, respectively, we prove in Sec.~\ref{subsec:descendantnorm} that
\begin{align} \label{eq:pseudo_orthog}
    {}_{\mu}\braket{0}{0}_{\mu'} & = \delta_{M,M'} \!\! \sum_{n_1 < \dots < n_M}^N \!\!\! \bigl| \Delta (\omega^{n_1}, \dots\mspace{-1mu}, \omega^{n_M}) \bigr|^{\mspace{1mu}4} \, \mathsf{P}_{\nu} (\omega^{-n_1}, \dots, \omega^{-n_M}) \,  \mathsf{P}_{\nu'} (\omega^{n_1} , \dots, \omega^{n_M} ) \nonumber \\[1ex]
    & \propto \ \delta_{\mu, \mu'} \; ,
\end{align}
with Jack polynomials depending on partitions $\nu = \nu (\mu) $ and $\nu' = \nu(\mu') $ obtained from the motifs as in \eqref{eq:nu_from_mu}. Moreover, we show that the \textsc{yhw} states have norm-squared given by the simple factorised formula \eqref{eq:normHigh}. This formula should also arise in the freezing limit of a special case of the norms found in \cite{TU}. Instead, we provide a simple and self-contained proof in Sec.~\ref{subsec:descendantnorm}.

\subsection{Yangian descendants}
\label{subsec:Ydescendant}

Even though all \textsc{yhw} states of the HS chain are known explicitly, various physical applications require one to construct the rest of each Yangian multiplet. Part of this is straightforward: we can apply the spin lowering operator $\mathbf{S}^- = \mathbf{S}^x - \I \, \mathbf{S}^y$ to any eigenvector to get a descendant for the $\mathfrak{su}_2$ (or really: $\mathfrak{sl}_2$) subalgebra of the Yangian. Thus we focus on the remaining Yangian descendants, which we will call `affine descendants'. In principle, one can apply the affine lowering operator $\mathbf{Q}^- = \mathbf{Q}^x - \I \, \mathbf{Q}^y$ to get such descendants. However, those are not necessarily orthogonal to the spin descendants or each other. For applications to correlation functions and quantum quenches, it is crucial to have an orthogonal eigenbasis for each Yangian multiplet.

The focus of this work is the systematic construction of such eigenstates forming an orthogonal eigenbasis along with the computation of their norms and overlaps. 

In the following we outline our results.

\paragraph{Strategy.} Following \cite{FLLS}, we use the method of the algebraic Bethe ansatz (\textsc{aba}) to construct affine descendants that are orthogonal to each other. The Yangian symmetry of the HS chain means that no Yangian generator is able to move between different eigenspaces. Thus, for each motif, we have an isolated eigenspace, with \textsc{yhw} state \eqref{eq:Y_hw}--\eqref{eq:Y_hw_wf}. Each of these Yangian multiplets can in turn be viewed as an inhomogeneous Heisenberg \textsc{xxx} chain with specific inhomogeneities that depend on the motif.\footnote{\ The values of the inhomogeneities are such that fusion occurs (often), cf.~\cite{FLLS}, and the Bethe ansatz is complete~\cite{Mukhin_2009, Mukhin_2014, Chernyak:2020lgw}.} This was called the `effective inhomogeneous spin chain' in \cite{FLLS}. It comes with a transfer matrix which, like the rest of the Yangian, commutes with the HS Hamiltonian; these are the `Heisenberg-style symmetries' of \cite{FLLS}. By simultaneously diagonalising them using the \textsc{aba}, we obtain an orthogonal eigenbasis of, in particular, the HS chain.

\paragraph{Results.} 
Inside each Yangian multiplet, Yangian descendant states can be constructed using the \textsc{aba}
\begin{equation} \label{eq:internalABA}
    \ket{\vect{u}}_{\mu} = \prod_{k=1}^{K} \! \mathbf{B}(u_k) \, \ket{0}_{\mu} \; ,
\end{equation}
with $\ket{0}_{\mu}$ the \textsc{yhw} state for the motif $\mu$ labelling the multiplet. Since any \textsc{yhw} state is an HS eigenstate and the Yangian is a symmetry, the `off-shell' Bethe vectors \eqref{eq:internalABA} are already eigenstates for the HS chain. However, they are not yet orthogonal.
For each Yangian multiplet, we obtain `effective' \textsc{bae} (or equivalently `effective' $TQ$-relations) for the the spectral parameters $\vect{u} = \{ u_k \}_{k=1}^K$ in order for \eqref{eq:internalABA} to furthermore become eigenstates of the transfer matrix. As long as these `on-shell Bethe vectors' all have different eigenvalues for the transfer matrix, they are guaranteed to be orthogonal.

In more detail, like in \cite{FLLS}, we work with the (diagonally) twisted transfer matrix $\mathbf{T}(x; \kappa) = \kappa \, \mathbf{A} (x) + \kappa^{-1} \, \mathbf{D} (x)$. The off-shell Bethe vectors are independent of the twist $\kappa$, but the effective \textsc{bae} do depend on $\kappa$. Three particular values of $\kappa$ are of special interest. First, we consider the isotropic limit $\kappa \to 1$. In this case, the Yangian descendants include the $\mathfrak{sl}_2$ descendants, which is useful for some physical applications. Second, $\kappa \to \infty$ or $\kappa \to 0$ is the so-called `Gelfand--Tsetlin (GT) limit', where the effective \textsc{bae} simplify tremendously and can be solved explicitly. The advantage of this limit is that we avoid having to solve non-trivial \textsc{bae}, while the Yangian descendant states obtained by acting with B-operators at those combinatorial Bethe roots remain non-trivial.

In addition, standard Bethe-ansatz technology allows us to compute the norms of the Yangian descendants, as well as the overlaps between on-shell and off-shell Bethe vectors within each HS eigenspace, in terms of the Gaudin and Slavnov determinants, respectively.

\section{Formalism: Dunkls, spin-Calogero--Sutherland and freezing}
\label{sec:CSbasis}

The exact solvability/quantum integrability of the Haldane--Shastry spin chain comes from its close relation to the trigonometric spin-Calogero--Sutherland (CS) system. The latter is a quantum-integrable system of $N$ spin-$1/2$ particles moving on a ring while interacting through long-range forces \cite{Ha:1992zz, Minahan:1992ht, Hikami_1993}. 
From it, the HS chain can be derived by taking a specific `freezing limit'. In this limit, the particle positions become fixed at equally spaced points on the ring, with residual interactions that are mediated solely through their spin degrees of freedom, thereby reducing to a spin chain. Thus, the HS chain can be viewed as the frozen, spin-only sector of the spin-CS system.

Adopting this perspective is advantageous, because the integrable structure of the spin-CS system, based on a framework using Dunkl operators, is inherited by the HS chain in the freezing limit~\cite{BGHP}. This connection allows us to leverage the full power of the spin-CS system's integrability, and in particular its underlying algebraic structure, to compute physical quantities such as norms and overlaps of eigenstates.

In this section, we provide a brief review of the spin-CS system and its integrability using Dunkl operators and the algebraic Bethe ansatz~(\textsc{aba}). We then consider the freezing limit and discuss the corresponding \textsc{aba} description for the HS chain. We follow \cite{FLLS,LS}, where more details can be found.

We will use tildes to denote the operators of the spin-CS model, to distinguish them from their counterparts for the HS chain, obtained in the freezing limit. 
We consider the fermionic spin-1/2 CS system. Its (total) momentum and Hamiltonian are~\cite{Ha:1992zz, Minahan:1992ht, Hikami_1993}
\begin{equation}
    \widetilde{\mathbf{P}}_{\textsc{cs}} = \sum_{i=1}^N \frac{\partial}{\partial x_i} \, , \qquad 
    \widetilde{\mathbf{H}}_{\textsc{cs}} = -\frac{1}{2} \sum_{i=1}^N \frac{\partial^2}{\partial x_i^2} + \sum_{i<j}^N \frac{\beta \, (\beta+ \mathbf{P}_{ij})}{4\sin^2[(x_i-x_j)/2]} \; ,
\end{equation}
where $x_i$ are the positions of the fermions,  $\mathbf{P}_{ij}$ is the spin permutation operator \eqref{eq:perm}, and $\beta$ is a (`reduced') coupling constant that will be sent to infinity in the freezing limit.

Define the complex multiplicative coordinates $z_j=e^{\mathrm{i}x_j}$. It is convenient to use a `gauge transformation' and define
\begin{equation} \label{eq:P' H'}
    \widetilde{\mathbf{P}}_{\textsc{cs}}' = \Phi_0(\bm{z})^{-1} \,\widetilde{\mathbf{P}}_{\textsc{cs}} \,\Phi_0(\bm{z}) , \qquad \widetilde{\mathbf{H}}^{\prime}_{\textsc{cs}} = \Phi_0(\bm{z})^{-1} \,\widetilde{\mathbf{H}}_{\textsc{cs}} \, \Phi_0(\bm{z}) \; ,
\end{equation}
where we conjugate by
\begin{equation} \label{eq:Phi0}
    \Phi_0(\bm{z})=\prod_{i\ne j}^N\biggl(1-\frac{z_i}{z_j}\biggr)^{\!\beta/2} \; .
\end{equation}

We will not require explicit expressions for the gauge-transformed operators~\eqref{eq:P' H'} for the remainder of this paper, but see e.g.\ \cite{LS}. Let $s_{ij}$ denote the operator that swaps $z_i$ and $z_j$.
The benefit of the gauge-transformed system is that it makes sense on a simple Hilbert space, namely the subspace of the space $\mathbb{C}[z_1^{\pm1} , \dots , z_N^{\pm1}] \otimes \left(\mathbb{C}^2 \right)^{\otimes N}$ of vector-valued Laurent polynomials given by the antisymmetry constraint 
\begin{equation} \label{eq:fermionic}
    \mathbf{P}_{i,i+1} \, s_{i,i+1} \, \ket{\widetilde{\psi}} = - \ket{\widetilde{\psi}} \; , \qquad i = 1,\dots, N-1 \; .
\end{equation}
This fermionic condition reflects the anticommutation when swapping fermions.

\paragraph{Dunkl operators.} 
The integrability of the spin-CS model is based on Dunkl operators, which will play the role of `dynamical inhomogeneities'. The (Cherednik--)Dunkl operators for a system of $N$ distinguishable particles are 
\begin{equation}
    d_i=\frac{1}{\beta} \, z_i \, \frac{\partial}{\partial z_i} - \sum_{j=1}^{i-1} \frac{z_i}{z_{ji}} \, (1-s_{ij}) +\! \sum_{j=i+1}^N \frac{z_j}{z_{ij}} (1-s_{ij}) \, + \frac{N+1-2\,i}{2} \; ,
    \label{eq:Dunklop}
\end{equation}
where we used the shorthand $z_{ij}\equiv z_i-z_j$. Since $z_i \, \partial_{z_i} = -\I \,\partial_{x_i}$ is the usual quantum-mechanical momentum operator in disguise, \eqref{eq:Dunklop} resembles a covariant derivative with terms exchanging particle positions. Although their definition might appear complicated, Dunkl operators possess elegant algebraic properties. Specifically, they satisfy the relations
\begin{equation} 
    d_i\,d_j=d_j\,d_i,\quad d_i\,s_{i,i+1}=s_{i,i+1}\,d_{i+1}+1,\quad d_i\,s_{jk}=s_{jk}\,d_i\quad\text{for}\quad i\neq j,k
    \label{eq:dAHA}
\end{equation}
of the degenerate affine Hecke algebra~\cite{Drinfeld:1986in}.

The spin-CS conserved quantities \eqref{eq:P' H'} are symmetric polynomials of Dunkl operators,
\begin{equation}
    \widetilde{\mathbf{P}}^{\prime}_{\textsc{cs}} = \beta \sum_{i=1}^N d_i \; , \quad \widetilde{\mathbf{H}}^{\prime}_{\textsc{cs}} = \frac{\beta^2}{2} \left( \, \sum_{i=1}^N d_i^2 \; - \widetilde{E}_0 \right) \; ,
\end{equation}
where $\widetilde{E}_0 = \sum_{i=1}^N (N-2\,i+1)^2/2 = N\,(N^2-1)/12$ is a constant.

\paragraph{Yang--Baxter integrability.} 
Now we can consider the Yang--Baxter integrability of the model. We start with the rational $R$-matrix on $\mathbb{C}^2 \otimes\mathbb{C}^2$ normalised as
\begin{align}
    \mathbf{R}(x)=x + \mathbf{P} \; ,
\end{align}
where $x$ is a spectral parameter. The $R$-matrix satisfies the Yang--Baxter equation,
\begin{equation}
    \mathbf{R}_{12}(x-y) \, \mathbf{R}_{13}(x) \, \mathbf{R}_{23}(y) = \mathbf{R}_{23}(y) \, \mathbf{R}_{13}(x) \, \mathbf{R}_{12}(x-y) \; ,
    \label{eq:YBE}
\end{equation}
on the vector space $\mathbb{C}^2 \otimes\mathbb{C}^2\otimes\mathbb{C}^2$. As in \eqref{eq:perm}, the subscripts specify on which two tensor factors each $R$-matrix acts nontrivially.

To construct the monodromy matrix, we follow the quantum inverse-scattering method and then replace the inhomogeneities $\in \mathbb{C}$ by the Dunkl operators $-d_j$ \cite{BGHP,Drinfeld:1986in} 
\begin{equation} \label{eq:defMono}
\begin{aligned}
    \widetilde{\mathbf{M}}_a (x) & = \mathbf{R}_{a1}\Bigl(x+d_1-\tfrac{1}{2}\Bigr)\cdots \mathbf{R}_{aN}\Bigl(x+d_N-\tfrac{1}{2}\Bigr) \\
    & = \Bigl(x+d_1-\frac{1}{2} +\mathbf{P}_{a1}\Bigr) \cdots \Bigl(x+d_N -\frac{1}{2} +\mathbf{P}_{aN}\Bigr) \,.
\end{aligned} 
\end{equation}
Here the subscript $a$ denotes the auxiliary space $\mathbb{C}^2$, while the subscripts $1,\dots,N$ enumerate the local spin-$1/2$ Hilbert spaces inside $(\mathbb{C}^2)^{\otimes N}$. Meanwhile, the Dunkl operators act non-trivially on polynomials. Since the Dunkls commute amongst themselves, see \eqref{eq:dAHA}, the standard proof based on the Yang--Baxter equation~\eqref{eq:YBE} shows that the monodromy matrix satisfies the RTT relation
\begin{equation} \label{eq:RMM}
    \mathbf{R}_{ab}(x-y) \, \widetilde{\mathbf{M}}_a (x) \, \widetilde{\mathbf{M}}_{b} (y)= \widetilde{\mathbf{M}}_{b}(y) \, \widetilde{\mathbf{M}}_{a}(x) \, \mathbf{R}_{ab}(x-y)  \; .
\end{equation}
A priori, it is not obvious that the monodromy matrix \eqref{eq:defMono} is moreover compatible with the fermionic constraint \eqref{eq:fermionic} of the Hilbert space. Thanks to the relations~\eqref{eq:dAHA} one can verify that, if $\ket{\widetilde{\psi}}$ is fermionic, then so is $\widetilde{\mathbf{M}}_a (x) \, \ket{\widetilde{\psi}}$, showing that \eqref{eq:defMono} does indeed act on the Hilbert space.\footnote{\ The sign of the Dunkl operators in \eqref{eq:defMono} depends on the order of the factors of the monodromy matrix and whether one considers the bosonic or fermionic spin-CS system. The monodromy matrix \eqref{eq:defMono} is ordered as in \cite{FLLS}, for which the fermionic constraint \eqref{eq:fermionic} requires replacing the usual inhomogeneities by $-d_j$. Ref.\ \cite{LS} uses the opposite monodromy matrix, in which case the signs of the Dunkl operators are opposite: the monodromy matrix $\mathbf{R}_{aN}(x-d_N-\tfrac{1}{2})\cdots \mathbf{R}_{a1}(x-d_1-\tfrac{1}{2})$ is also compatible with \eqref{eq:fermionic}.}

It is possible to extract from this monodromy matrix fully explicit generators in Drinfeld's first presentation (``Drinfeld $J$-presentation'') of the Yangian \cite{Bernard:1994wg,LS}. For our purposes, however, the above RTT presentation is more convenient. As usual, the monodromy matrix can be written as a $2\times 2$ matrix on the auxiliary space
\begin{equation}
    \widetilde{\mathbf{M}}_a (x)=\begin{pmatrix} \widetilde{\mathbf{A}}(x) & \widetilde{\mathbf{B}}(x) \\ \widetilde{\mathbf{C}}(x) & \widetilde{\mathbf{D}}(x)  \end{pmatrix}_{\!a} \; ,
\end{equation}
where the four matrix entries are `quantum' operators that again act on the fermionic Hilbert space. To obtain the Hamiltonian and the higher conserved charges of the CS system, we consider the quantum determinant
\begin{equation}
    \mathrm{qdet}_a \, \widetilde{\mathbf{M}}_a (x) = \widetilde{\mathbf{A}}(x) \, \widetilde{\mathbf{D}}(x-1)- \widetilde{\mathbf{B}}(x)\, \widetilde{\mathbf{C}}(x-1) \;.
\end{equation}
The conserved charges can be obtained by expanding the following combination
\begin{align} \label{eq:Delta(x)}
    \widetilde{\Delta}(x) & = \frac{\mathrm{qdet}_a \, \widetilde{\mathbf{M}}_a (x)}{\mathrm{qdet}_a \, \widetilde{\mathbf{M}}_a (x-1)} = \prod_{i=1}^N \frac{x-d_i+\frac{1}{2}}{x-d_i-\frac{1}{2}} \\ 
    & = \,1 + N \, x^{-1} + \left(\frac{N^2}{2}+\frac{\widetilde{\mathbf{P}}'_\textsc{cs}}{\beta}\right)x^{-2}+\left(\frac{N^3}{4}+\frac{N\,\widetilde{\mathbf{P}}'_\textsc{cs}}{\beta}+\frac{2\,\widetilde{\mathbf{H}}'_\textsc{cs}}{\beta^2}\right)x^{-3}+\mathcal{O}(x^{-4})\; , \nonumber
\end{align}
where $N$ is the particle number, and
$\widetilde{\mathbf{P}}'_\textsc{cs}$ and $\widetilde{\mathbf{H}}'_\textsc{cs}$ are operators in \eqref{eq:P' H'}. The coefficients of higher-order terms in the $x^{-1}$-expansion contain higher conserved charges.
 
Indeed, since the Dunkl operators commute amongst each other, we have
\begin{equation}
    [\widetilde{\Delta}(x),\widetilde{\Delta}(y)] = 0 \; ,
\end{equation}
so that all coefficients in the $x^{-1}$-expansion commute with each other. The quantum determinant generates the center of the Yangian $Y(\mathfrak{gl}_2)$, the algebra generated by the quantum operators $\widetilde{\mathbf{A}}(x)$, $\widetilde{\mathbf{B}}(x)$, $\widetilde{\mathbf{C}}(x)$ and $\widetilde{\mathbf{D}}(x)$. In particular, it means that the quantum determinant commutes with all elements of the monodromy matrix:
\begin{equation}
    [\widetilde{\Delta}(x), \widetilde{\mathbf{A}}(y)]=[\widetilde{\Delta}(x), \widetilde{\mathbf{B}}(y)]=[\widetilde{\Delta}(x), \widetilde{\mathbf{C}}(y)]=[\widetilde{\Delta}(x), \widetilde{\mathbf{D}}(y)]=0\,.
\end{equation}
This further implies that the higher conserved charges have Yangian symmetry.

It is crucial to emphasize the difference between the integrable structure of the spin-CS system and that of the Heisenberg chain. In the latter, the Hamiltonian and higher charges are contained in the transfer matrix $\widetilde{\mathbf{T}}(x)=\mathrm{tr}_a \, \widetilde{\mathbf{M}}_a (x) = \widetilde{\mathbf{A}}(x) + \widetilde{\mathbf{D}}(x)$, which satisfies $[\widetilde{\mathbf{T}}(x), \widetilde{\mathbf{T}}(y)]=0$ but does not commute with the A-, B-, C-, D-operators individually. Instead, the latter are used in the algebraic Bethe ansatz. In contrast, the Hamiltonian of the CS model (and consequently the HS chain), and the higher charges contained in $\widetilde{\Delta}(x)$, commute with the entire Yangian. If one further defines transfer matrices for the spin-CS system, one obtains additional conserved charges that commute with $\widetilde{\Delta}(x)$ but do not have the Yangian symmetry~\cite{FLLS}. We will return to this point in the next section.

\paragraph{Eigenspaces and inhomogeneous Heisenberg chains.}
The definition of the monodromy matrix in \eqref{eq:defMono} is somewhat formal because $d_j$ are themselves differential operators. 
When acting on eigenstates, we can replace the Dunkl operators by their eigenvalues.

The Dunkl operators are simultaneously diagonalized by the so-called nonsymmetric Jack polynomials $E_{\bm{\lambda}}(\bm{z})=E_{\bm{\lambda}}^{(1/\beta)}(\bm{z})$ (see \cite{LS, FLLS} for more details and further references), where $1/\beta$ is the Jack parameter. For a given set of `quantum numbers' $\bm{\lambda}=(\lambda_1,\ldots,\lambda_N)\in\mathbb{Z}^N$, the nonsymmetric Jack polynomial $E_{\bm{\lambda}}(\bm{z})$ is uniquely characterized by the conditions
\begin{equation}
    E_{\bm{\lambda}}(\bm{z})=z_1^{\lambda_1}\cdots z_N^{\lambda_N}+\text{lower} \, , \qquad d_i\,E_{\bm{\lambda}}(\bm{z})=\delta_i(\bm{\lambda}) \, E_{\bm{\lambda}}(\bm{z}) \, , 
\end{equation}
where `lower' refers to a partial ordering on monomials called the `dominance order', see e.g.~\cite{LS}. The eigenvalues $\delta_i(\bm{\lambda})$ depend only on the partition $\bm{\lambda}^+$ obtained by sorting the parts of $\bm{\lambda}$ into weakly decreasing order. For example, if $\bm{\lambda} = (3,-1,0,0,5,0)$ then $\bm{\lambda}^+ = (5,3,0,0,0,-1)$. For the fermionic spin-CS system, only strict partitions (where all $\lambda_i^+$ are distinct) are allowed, and such strictly decreasing sequences of integers label the eigenspaces of the fermionic spin-CS system, with exact wavefunctions expressed in terms of nonsymmetric Jacks \cite{TU}. Moreover, these eigenspaces are precisely irreducible representations of the Yangian.

For $\bm{\lambda} = \bm{\lambda}^+$ a partition, the eigenvalues of the Dunkl operators are
\begin{equation} \label{eq:eigenDunkl}
    \delta_i(\bm{\lambda})=\frac{\lambda_i}{\beta}+\frac{1}{2}\,(N+1-2\,i),\qquad i=1,\ldots,N \,.
\end{equation}
When we restrict the monodromy matrix to an eigenspace of the spin-CS system, we may replace the Dunkl operators by their eigenvalues. For the eigenspace labelled by the strict partition $\bm{\lambda}$, the monodromy matrix reads 
\begin{equation}
    \widetilde{\mathbf{M}}_a(x)\big|_{\bm{\lambda}} =\mathbf{R}_{a1}\Bigl(x+\delta_1(\bm{\lambda})-\tfrac{1}{2}\Bigr) \cdots  \mathbf{R}_{aN}\Bigl(x+\delta_N(\bm{\lambda})-\tfrac{1}{2}\Bigr) \, .
\end{equation}
This is nothing but the monodromy matrix of an inhomogeneous Heisenberg XXX spin chain with particular values for the inhomogeneities.

\paragraph{Freezing limit.} 
As alluded to previously, the HS chain is obtained by taking a `freezing limit' of the spin-CS system \cite{Polychronakos:1992ki,BGHP,Talstra_1995}. A rigorous treatment uses deformation quantization, see \cite{chalykh24,Liashyk:2024ekz}; here we suffice with the physical picture, cf.~e.g.~\cite{LS}. The freezing limit can be viewed as a strong coupling limit $\beta\to\infty$ supplemented by evaluating the coordinates $z_j$ at distinct $N$th roots of unity,
\begin{equation} \label{eq:evaluationmap}
    \text{ev}\colon (z_1,\ldots,z_N) \to (1,\omega,\omega^2,\dots,\omega^{N-1}) \, , \qquad \omega^n = \mathrm{e}^{2\mathrm{i}\pi n/N}\; .
\end{equation}
Under this combined limit, the spin-CS Hamiltonian reduces to the Haldane--Shastry spin chain
\begin{equation}
    \beta^{-1}\,\widetilde{\mathbf{H}}'_\textsc{cs} \;\to\; \mathbf{H}_{\textsc{hs}}=\sum_{i<j}^N \frac{1+\mathbf{P}_{ij}}{4\sin^2\left[\frac{\pi}{N}(i-j)\right]}\,,
\end{equation}
which is the antiferromagnetic version of \eqref{eq:H_HS}. 

Like the HS Hamiltonian, its higher-order charges are directly obtained from the corresponding operators in the spin-CS system via the freezing limit. 

The Yangian symmetry is also directly inherited from freezing \cite{Bernard:1994wg,LS}. 
For instance, as mentioned above, from the monodromy matrix \eqref{eq:defMono} one can extract the $\mathfrak{su}_2$ generators \eqref{eq:Salpha} along with operators $\widetilde{\mathbf{Q}}^\alpha$ of the Drinfeld $J$-presentation of the Yangian. Using the fermionic condition~\eqref{eq:fermionic} one can replace the coordinate permutations $s_{ij}$ in the Dunkl operators by $-\mathbf{P}_{ij}$ when they are on the right in the expression, cf. \eqref{eq:fermionic}. In the limit $\beta\to\infty$, the derivatives of the Dunkl operators drop out, and the evaluation of the remaining rational functions of the coordinates $z_j$ finally gives rise to trigonometric functions. This is how \eqref{eq:Qalpha} is obtained, see App.~C of \cite{LS} for details. 

For our purposes, the quantum operators provide a more useful set of generators of the Yangian. In principle, one can go through the same steps as above, and define
\begin{equation} \label{eq:mono_freezing}
    \mathbf{M}_a(x) \coloneqq \mathrm{ev}\Bigl(\, \lim_{\beta\to\infty} \widetilde{\mathbf{M}}_a(x) \Bigr) \, .
\end{equation}
Unfortunately, we do not know how to express the result in terms of spin operators only in a conceptual or useful way. Nevertheless, as we explain below, one can work with \eqref{eq:mono_freezing} by remembering its origin before freezing.

In the limit the partitions $\bm{\lambda}$ are replaced by motifs, see~\cite{LS}. However, we do not know a similarly direct way to get the eigenstates of the HS chain from eigenstates of the spin CS system by freezing, see also \textsection 3.2 of \cite{LS}. While still based on freezing (whence the $\omega^n$), the derivation of the \textsc{yhw} wave functions \eqref{eq:Y_hw_wf} is more indirect \cite{BGHP,LS}.

\paragraph{Quantum operators for HS.} 
For the \textsc{aba}, we need to understand how to work with the quantum operators $\mathbf{A}(x), \mathbf{B}(x), \mathbf{C}(x),\mathbf{D}(x)$ contained in \eqref{eq:mono_freezing} on the spin-chain Hilbert space $\left( \mathbb{C}^2 \right)^{\otimes N}$. 
For the HS chain, these are inherited from the spin-CS system via freezing, which means that working with them requires a little more effort than for usual Heisenberg chains.

We start with the action on an arbitrary \textsc{yhw} state $\ket{0}_\mu$. The formula \eqref{eq:Y_hw} can be re-expressed as 
\begin{equation}  \label{eq:Y_hw_v2}
    \ket{0}_\mu = \sum_{n_1 < \dots < n_M}^N \!\!\!\!\! \mathrm{ev}\Bigl( \, \widetilde{\Psi}_\mu(z_{n_1},\dots,z_{n_M}) \Bigr) \, \cket{n_1,\dots,n_M} \, , 
\end{equation}
where `ev' denotes the `evaluation' map~\eqref{eq:evaluationmap}, and the polynomial 
\begin{equation} \label{eq:Y_hw_wf_v2}
    \widetilde{\Psi}_\mu(z_1,\dots,z_M) = z_1 \cdots z_M \, V(z_1, \dots, z_M)^2 \, \mathsf{P}_{\nu(\mu)} (z_1 , \dots, z_M) 
\end{equation}
evaluates to \eqref{eq:Y_hw}.

The action of elements of the monodromy matrix on HS states such as \eqref{eq:Y_hw_v2} should be understood as
\begin{equation} \label{eq:Y_HS_via_spinCS}
    \mathbf{M}_a(x) \, \ket{0}_\mu \coloneqq \sum_{n_1  < \cdots < n_M }^N \!\!\!\!\! \mathrm{ev} \biggl(\, \lim_{\beta \to \infty} \widetilde{\mathbf{M}}_a(x) \ \widetilde{\Psi}_\mu(z_{n_1},\dots,z_{n_M}) \biggr) \, \cket{n_1,\dots,n_M} \, .
\end{equation}
By taking matrix entries in the auxiliary space, one gets analogous formulas for the quantum operators $\mathbf{A}(x)$, $\mathbf{B}(x)$, $\mathbf{C}(x)$ and $\mathbf{D}(x)$. 

Broken down into steps, \eqref{eq:Y_HS_via_spinCS} gives the following recipe for working with a Yangian operator $\mathbf{X}$ of the HS chain:
\begin{enumerate}
    \item Take the corresponding spin-CS Yangian operator $\widetilde{\mathbf{X}}$.
    \item Omit the derivatives in all Dunkl operators $d_j$ \eqref{eq:Dunklop} contained in $\widetilde{\mathbf{X}}$ (since $\beta\to\infty$). 
    \item Write the wavefunction of the HS eigenstate of interest in the form $\mathrm{ev}\bigl( \widetilde{\Psi}(\vect{z}) \bigr)$ where $\widetilde{\Psi}(\vect{z})$ is a suitable polynomial, e.g.\ \eqref{eq:Y_hw_wf_v2} in the case of $\ket{0}_\mu$. \label{it:HS_wf_from_ev}
    \item Act with the swaps $s_{ij}$ in the Dunkl operators in $\widetilde{\mathbf{X}}$ on the variables $z_i$ of $\widetilde{\Psi}(\vect{z})$ to get a vector with entries that are polynomials. 
    \item Evaluate the $z_i$ in this result to numbers to obtain a state in the HS Hilbert space.
\end{enumerate}
It can be shown that step~\ref{it:HS_wf_from_ev} is always possible in principle.
\footnote{\ In short, the argument goes as follows. Step~\ref{it:HS_wf_from_ev} holds for the \textsc{yhw} states, see \eqref{eq:Y_hw_v2}--\eqref{eq:Y_hw_wf_v2}. The unevaluated vector can in fact be lifted to an element of the spin-CS Hilbert space. (This is nontrivial.) By completeness, any other vector can be obtained from \textsc{yhw} states using the HS Yangian. But the latter come from spin-CS Yangian operators, which send fermionic spin-CS vectors to fermionic spin-CS vectors. Thus, Step~\ref{it:HS_wf_from_ev} holds for any eigenstate.}
For our purposes, we only need the polynomial \eqref{eq:Y_hw_wf_v2}, since we will only have to apply quantum operators to the \textsc{yhw} states $\ket{0}_\mu$ as in \eqref{eq:Y_HS_via_spinCS} to carry out the \textsc{aba}.

\section{Bethe vectors for the Haldane--Shastry chain}
\label{sec:HSABA}

\subsection{Internal Bethe ansatz}
\label{subsec:textsc{aba}}

Following \cite{FLLS}, all eigenstates of the HS chain can be constructed systematically using the algebraic Bethe ansatz (\textsc{aba}).

\paragraph{Yangian highest-weight states}
The starting point of the \textsc{aba} is a \textsc{yhw} state. Unlike for the Heisenberg chain, where all states can be reached starting from the ferromagnetic state $\ket{\uparrow\cdots\uparrow}$ acting by B-operators, for HS there are multiple \textsc{yhw} states $\ket{0}_{\mu}$: one for each motif $\mu$. These are the states \eqref{eq:Y_hw} introduced in Sec.~\ref{subsec:YHW}, with wave functions \eqref{eq:Y_hw_wf} containing (a special case of) Jack polynomials.

By definition, a \emph{Yangian highest-weight (\textsc{yhw}) state}, or \emph{pseudovacuum}, is an eigenvector of both the A- and D-operators that is annihilated by the C-operator,
\begin{equation} \label{eq:pseudovacuum}
    \begin{pmatrix}
        \mathbf{A}(x) & \mathbf{B}(x) \\ \mathbf{C}(x) & \mathbf{D}(x)
    \end{pmatrix}_{\!a} \ket{0}_\mu = 
    \begin{pmatrix}
        A_\mu(x) \, \ket{0}_\mu & \ast \\ 0 & D_\mu(x) \, \ket{0}_\mu 
    \end{pmatrix}_{\!a} \, ,
\end{equation}
where `$\ast$' denotes a certain vector. 
This is the appropriate notion of `highest-weight vector' for the Yangian, generalising the familiar conditions $S^+ \, \ket{\Psi} = 0$ and $S^z \, \ket{\Psi} = s\,\ket{\Psi}$ for the Lie algebra $\mathfrak{sl}_2$. The corresponding analogue of its highest weight (spin-$s$ in $\mathfrak{sl}_2$ case) is the \emph{Drinfeld polynomial} $P_\mu(x) = x^N + \sum_{n=0}^{N-1} c_n \, x^n$, which is determined by the eigenvalues $A_\mu(x)$ and $D_\mu(x)$ through the relation
\begin{equation} \label{eq:A_over_D_vs_Dri}
    \frac{A_\mu (x)}{D_\mu (x)} = \frac{P_\mu (x+1)}{P_\mu (x)} \; .
\end{equation}
The ratio avoids any dependence on common rescalings of the (R-matrix and thus) A- and D-operators. 
For $\ket{0}_\mu$, it can be shown that the eigenvalues of the A- and D-operators in the freezing limit are polynomials of degree $N$ with zeroes at half-integers~\cite{BGHP,Bernard:1994wg}, namely
\begin{align}
    A_\mu(x) & = \prod_{n=0}^{N-1} \biggl( x + \frac{N+1-2\,n}{2} \biggr) \prod_{n\in \mu} \frac{x + \frac{N-2\,n-1}{2}}{x + \frac{N-2\,n+1}{2}} \notag\\
    & = \prod_{n =0}^{N-1} \biggl( x + \frac{N+1-2\,(n + \delta_{n \in \mu})}{2} \biggr) \, , \label{eq:Aeigenvaluehw} \\
    D_{\mu}(x) & = \prod_{n=1}^{N} \biggl( x + \frac{N+1-2\,n}{2} \biggr) \prod_{n\in \mu} \frac{x + \frac{N-2\,(n-1)+1}{2}}{x + \frac{N-2\,(n-1)-1}{2}} \notag\\
    & = \prod_{n =1}^{N} \biggl( x + \frac{N+1-2\,(n - \delta_{n \in \mu})}{2} \biggr) \, .
    \label{eq:Deigenvaluehw}
\end{align}
where $\delta_{n\in \mu}=1$ if $n\in\mu$ and $\delta_{n\in \mu}=0$ else. This yields the Drinfeld polynomial \eqref{eq:Drinfeldpoly}.

\paragraph{Off-shell Bethe vectors.} 
From the \textsc{yhw} state $\ket{0}_\mu$ one obtains Yangian descendant states, which belong to the same motif $\mu \in \mathcal{M}_N$, by acting by B-operators with rapidities $\vect{u} = \{u_k\}_{k=1}^K$ as usual:
\begin{equation} \label{eq:internalABAoff}
    \ket{\vect{u}}_{\mu} \coloneqq \prod_{k=1}^{K} \! \mathbf{B}(u_k) \, \ket{0}_{\mu} \; .
\end{equation}
This is the off-shell Bethe vector \eqref{eq:internalABAoff}.
Note that it has $M+K$ spins $\downarrow$ if $\mu = (\mu_1,\dots,\mu_M)$. 

A special feature of the HS chain is that, since the HS Hamiltonian \emph{commutes} with the B-operator, all Yangian descendants are degenerate for the HS Hamiltonian,
\begin{equation} \label{eq:HS_offshell}
    \mathbf{H} \, \ket{\vect{u}}_{\mu} = E(\mu) \, \ket{\vect{u}}_{\mu} \; ,
\end{equation}
with the same HS energy \eqref{eq:EP}--\eqref{eq:disperseRel} for the entire Yangian multiplet. The same is true for the eigenvalues of all other conserved charges coming from the expansion of the quantum determinant.
We emphasise that the eigenvalue equation \eqref{eq:HS_offshell} holds for off-shell Bethe vectors, i.e.\ for \emph{arbitrary} values of the spectral parameters $\vect{u}$.

\paragraph{Transfer matrix.}
To get an eigen\emph{basis} from the off-shell Bethe vectors~\eqref{eq:internalABA} we need to fix the values of the spectral parameters $\vect{u}$ such that we get orthogonal vectors that span the eigenspace. It is usually preferable to choose an orthonormal basis. A natural way, inspired by the usual \textsc{aba} construction, is to require the Bethe vectors to be eigenstates of a transfer matrix $\mathbf{T}(x)$. Indeed, on the one hand, $\mathbf{T}(x)$,
which generates the so-called Bethe subalgebra of the Yangian, 
commutes with the HS Hamiltonian, and the two can be simultaneously diagonalised. On the other hand, as usual, $\mathbf{T}(x)$ does \emph{not} commute with $\mathbf{B}(y)$, so the Bethe vector $\ket{\vect{u}}_{\mu}$ is in general not an eigenstate of $\mathbf{T}(x)$, unless the rapidities $\vect{u}$ satisfy certain quantization conditions, which are nothing but the Bethe ansatz equations. It remains to choose a suitable transfer matrix.

Like in \cite{FLLS}, we will work with the (diagonally) \emph{twisted} transfer matrix, defined by
\begin{equation}
    \mathbf{T}(x;\kappa) \coloneqq \text{Tr}_a \pigl(\kappa^{\sigma_a^z} \, \mathbf{M}_a(x)\pigr) = \kappa\,\mathbf{A}(x)+\kappa^{-1} \, \mathbf{D}(x)\,,
    \label{eq:Ttwist}
\end{equation}
where the twist $\kappa\in \mathbb{C}$ is an arbitrary constant. 
From \eqref{eq:Ttwist} one can extract commuting charges as usual. These are called `Heisenberg-style charges' in \cite{FLLS}, in which some explicit examples were obtained, including \eqref{eq:Heis_style}--\eqref{eq:Heis_style_isotr}.

From \cite{TU} it follows\,%
\footnote{\ This property is more easily proved in the language of spin-CS system, where there exists an inner product (with measure the square of \eqref{eq:Phi0}) on functions of the coordinates such that the Dunkl operators are hermitian. This inner product extends to an inner product on $\mathbb{C}[z_1^{\pm1} , \dots , z_N^{\pm1}] \otimes (\mathbb{C}^2 )^{\otimes N}$ for which the monodromy matrix $\widetilde{\mathbf{M}}_a (x)$ is `hermitian' (self-adjoint) \cite[Prop.\,9]{TU}. In the freezing limit this implies \eqref{eq:ADconjugate}.}
that the quantum operators satisfy
\begin{equation}
    \mathbf{A}(x)^\dagger = \mathbf{A}(x^*) , \quad \mathbf{D}(x)^\dagger = \mathbf{D}(x^*) \; , \qquad x \in \mathbb{C} \; ,
    \label{eq:ADconjugate}
\end{equation}
where $x^*$ is the complex conjugate of $x$. Therefore, the twisted transfer matrix obeys
\begin{equation}
    \mathbf{T}(x;\kappa)^\dagger = \mathbf{T}(x^*;\kappa^*) \; .
\end{equation}
That is, the coefficients of the twisted transfer matrix as a formal power series in the spectral parameter $x$, including the Heisenberg-type charge \eqref{eq:Heis_style_isotr}, are hermitian when $\kappa$ is real.

\begin{quote}
    \emph{From now on, we will assume that the twist $\kappa$ is real, and take $\kappa>0$.} 
\end{quote}
The resulting hermiticity of the coefficients of  twisted transfer matrix as a formal power series in $x$ thus guarantees that its eigenstates will be orthogonal, as desired.

\paragraph{On-shell Bethe vectors.} 
The diagonalization of the twisted transfer matrix $\mathbf{T}(x;\kappa)$ is achieved using the Yangian relations \eqref{eq:RMM} in the same way as for the Heisenberg chain. To state the result, we consider Baxter's $Q$-polynomial
\begin{equation} \label{eq:Q}
    Q_{\mu,\vect{u}}(x) = \prod_{k=1}^K (x-u_k) \; ,
\end{equation}
whose zeroes are the spectral parameters $\vect{u}$ for a given motif $\mu$. 
The off-shell Bethe vector~\eqref{eq:internalABA} becomes an eigenstate of the transfer matrix,
\begin{equation}
    \mathbf{T}(x;\kappa) \, \ket{\vect{u}}_{\mu} = T_{\mu,\vect{u}}(x;\kappa) \, \ket{\vect{u}}_{\mu} \, ,
\end{equation}
with eigenvalue a polynomial in $x$ (despite the denominator in the following) of degree $N$,
\begin{equation} \label{eq:T_eigenvalue}
    T_{\mu,\vect{u}}(x;\kappa) = \frac{\kappa \, A_\mu (x) \, Q_{\mu,\vect{u}}(x-1) + \kappa^{-1} \, D_\mu (x)\, Q_{\mu,\vect{u}}(x+1)}{Q_{\mu,\vect{u}}(x)} \; ,
\end{equation}
provided the rapidities (Bethe roots) $\vect{u}$ satisfy the Bethe-ansatz equations (\textsc{bae})
\begin{equation}
    \kappa^2 \, \frac{A_\mu(u_k)}{D_\mu (u_k)} \, \frac{Q_{\mu,\vect{u}}(u_k -1)}{Q_{\mu,\vect{u}}(u_k +1)}= -1 \; , \qquad k = 1,\dots, K \; .
    \label{eq:effectiveBAE}
\end{equation}
Through these equations, the Bethe roots $\vect{u}$, and therefore the on-shell Bethe vectors, depend on the value of $\kappa$. 
Different choices of $\kappa$ thus lead to different orthogonal eigenbases. 

A more conventional rephrasing of the \textsc{bae} is
\begin{equation} \label{eq:BAE_v2}
    \kappa^2 \, \frac{A_\mu(u_k)}{D_\mu (u_k)} = \prod_{k' (\neq k)}^K \frac{u_k - u_{k'} +1}{u_k - u_{k'} -1} \; ,
\end{equation}
where the right-hand side is a product of the usual  `scattering phase' between two magnons for models based on a rational $R$-matrix, such as the Heisenberg \textsc{xxx} chain.

We observe that the familiar \textsc{bae} of the (twisted) Heisenberg \textsc{xxx} chain arise in the special case of the empty motif, where $\ket{0}_{\mu=\varnothing} = \ket{\uparrow\uparrow \cdots \uparrow}$ is the standard ferromagnetic \textsc{yhw} state of the Heisenberg chain, and \eqref{eq:Aeigenvaluehw}--\eqref{eq:Deigenvaluehw} become the usual eigenvalues of the A- and D-operators. More generally, a similar set of \textsc{bae} appear in the inhomogeneous Heisenberg \textsc{xxx} chain with linearly increasing inhomogeneities, see App.~\ref{app:inhomogeneousXXX}. 

We remark that just by solving the \textsc{bae} \eqref{eq:effectiveBAE}, or equivalently the $TQ$-relation 
\begin{equation}
    T_{\mu,\vect{u}} (x;\kappa) \, Q_{\mu,\vect{u}}(x) = \kappa \, A_\mu (x) \, Q_{\mu,\vect{u}}(x-1) + \kappa^{-1} D_\mu (x) \, Q_{\mu,\vect{u}}(x+1) \; .
    \label{eq:TQeffective}
\end{equation}
we find more solutions than what we expect from the representation theory. The reason is that there are \emph{unphysical} solutions if one just solves the \textsc{bae} \eqref{eq:effectiveBAE} or $TQ$-relation \eqref{eq:TQeffective}. Instead, we solve the Q-system of the inhomogeneous Heisenberg \textsc{xxx} chain with linearly increasing inhomogeneities, which gives only physical solutions, matching the representation theory. We explain the details in App.~\ref{app:Qsystem}. 

In the picture of the `effective' inhomogeneous Heisenberg \textsc{xxx} chain, the \textsc{yhw} states simply have singlets at the sites (in real space) indicated by the motif~\cite{FLLS}. More details on this `effective' inhomogeneous spin chain in the context of the HS chain (and spin-CS system) can be found in \cite[App.~B.1]{FLLS}.

\subsection{Periodic limit and \texorpdfstring{$\mathfrak{sl}_2$}{sl2}-descendants}

For the periodic case $\kappa = 1$, the transfer matrix does not just commute with $\mathbf{S}^z$, but has full $\mathfrak{sl}_2$ spin symmetry. The \textsc{bae} \eqref{eq:BAE_v2} simplify only slightly.

The $\mathfrak{sl}_2$-descendants are obtained by acting the global spin lowering operator $\mathbf{S}^- \sim \mathbf{B}(\infty)$ on all Bethe vectors with $\mathbf{S}^+ \, \ket{\vect{u}}_\mu = 0$. In terms of Bethe roots, the $\mathfrak{sl}_2$-descendants have the same finite Bethe roots as the corresponding $\mathfrak{sl}_2$ highest-weight state, with additional Bethe roots at infinity allowed in the limit $\kappa\to1$.

\subsection{Extreme twist and Gelfand--Tsetlin basis} \label{sec:GT}

Consider the normalized twisted transfer matrix
\begin{equation}
    \mathbf{t}(x ; \kappa) = \frac{1}{\kappa + \kappa^{-1}} \, \mathbf{T} (x ; \kappa) = \frac{\kappa}{\kappa + \kappa^{-1}} \, \mathbf{A}(x) + \frac{\kappa^{-1}}{\kappa + \kappa^{-1}} \, \mathbf{D} (x) \; .
    \label{eq:Ttwistnormalized}
\end{equation}
In the regime $\kappa \to \infty$ (or $\kappa \to 0$) of extreme twist, we are effectively diagonalising the quantum operator $\mathbf{A}(x)$ (or $\mathbf{D}(x)$, respectively). In terms of representation theory, either operator generates a $Y(\mathfrak{gl}_1)$ subalgebra of the full Yangian $Y(\mathfrak{gl}_2)$. It can be diagonalised simultaneously with the quantum determinant, \textit{i.e.}\ the HS Hamiltonian and the other `basic' conserved quantities. In the representation-theory literature, such an eigenbasis is known as a Gelfand--Tsetlin~(GT) basis \cite{Nazarov_1994,Molev_1994}. It is reproduced from the \textsc{aba} in the limit of extreme twist, where the \textsc{bae} \eqref{eq:BAE_v2} simplify drastically, and become `non-interacting', with explicit, purely combinatorial Bethe roots. 

For definiteness, consider the case $\kappa\to \infty$. The $TQ$-relations \eqref{eq:TQeffective} become
\begin{equation}
\label{eq:GTAeigenvalues}
    A_{\mu,\vect{u} } (x) \, Q_{\mu,\vect{u}} (x) = A_\mu (x) \, Q_{\mu,\vect{u}} (x-1) \; ,
\end{equation}
where $A_{\mu,\vect{u}} (x)$ denotes the eigenvalue of the A-operator on the off-shell Bethe vector $\ket{\vect{u}}_{\mu}$ while $A_\mu (x)$ is its eigenvalue \eqref{eq:Aeigenvaluehw} on the \textsc{yhw} state $\ket{0}_{\mu}$.

Recall from Sec.~\ref{subsec:yangian} the notion of `strings' for the zeroes of a Drinfeld polynomial. Like there, for a motif $\mu = (\mu_1,\dots,\mu_M)$ we set $\mu_0 \coloneqq -1$ and $\mu_{M+1} \coloneqq N+1$. The zeroes of the corresponding Drinfeld polynomial~\eqref{eq:Drinfeldpoly} consist of $M+1$ strings $\{ \mu_{m} +2-\frac{N+1}{2} , \dots , \mu_{m+1}-1-\frac{N+1}{2} \}$, for $0\leqslant m \leqslant M$, of half-integers (integer if $N$ is odd, half-odd else). Note that these strings have $\ell_m = \mu_{m+1} - \mu_{m} - 2$.

The Bethe roots at $\kappa\to\infty$ now simply consist of subsets of consecutive zeroes shifted by 1~\cite{Molev_1994}:
\begin{equation}
    \vect{u} = \vect{u}^{(\vect{\zeta})} \coloneqq \bigcup_{0 \leqslant m \leqslant M} \left\{ \mu_m + 1 - \frac{N+1}{2},\,\mu_m + 2 - \frac{N+1}{2},\,\dots,\,\mu_m + \zeta_m - \frac{N+1}{2}\right\} \; ,
\end{equation}
only depending on a choice of $M+1$ integers $0\leqslant \zeta_m \leqslant \mu_{m+1} - \mu_{m} - 2$ that indicate how many consecutive elements from the $m$th string of zeroes are included, starting from the left.\,%
\footnote{\ We remark that the same combinatorial structure appears for the $q$-deformed HS chain in the crystal limit $q \to 0$ \cite[\textsection1.2.5]{LPS}, where it is realised directly on the lattice: the affine highest-weight vectors become proportional to $\cket{\mu} = \prod_{n\in \mu} \sigma^-_n \; \ket{\uparrow\cdots\uparrow}$, and descendants correspond to extending domains of $\downarrow$s to the right without merging domains.} 
When $\zeta_0=\zeta_1 = \dots = \zeta_M = 0$ the set of Bethe roots is empty. 
In terms of these explicit Bethe roots, the GT eigenbasis at $\kappa\to\infty$ can be constructed by renormalising the Bethe vectors,
\begin{equation} \label{eq:GT_from_ABA}
    \ket{\textsc{gt}}_{\mu,\vect{u}} = \lim_{\kappa\to\infty} \Biggl(   \prod_{m=0}^{M} \, \prod_{k_m = 1}^{\zeta_m} \!\! \mathbf{B}^\mathrm{red} \biggl(\mu_{m} + k_m -\frac{N+1}{2}\biggr) \ \ket{0}_{\mu} \Biggl) \; ,
\end{equation}
where we use a `reduced' B-operator (see App.~\ref{app:reducedYangian}) to avoid a vanishing limit.

The corresponding $Q$-polynomial is
\begin{equation}
    Q_{\mu,\vect{u}}^{\textsc{gt}}(x) = \prod_{m=0}^{M} \; \prod_{k_m=1}^{\zeta_m} \!\! \left( x - \mu_m - k_m + \frac{N+1}{2} \right) \; .
\end{equation}
The corresponding eigenvalue of the A-operator reads 
\begin{equation}
    A_{\mu,\vect{u}} (x) = \prod_{n=0}^{N-1} \! \left( x + \frac{N+1-2\,n}{2} \right) \prod_{m=1}^M \frac{x+\frac{N-1-2\,\mu_m}{2}}{x+\frac{N+1-2\,\mu_m}{2}} \prod_{m'=0}^{M} \prod_{k_{m'}=1}^{\zeta_{m'}} \frac{x+\frac{N-1-\mu_{m'}-k_{m'}}{2}}{x+\frac{N+1-\mu_{m'}-k_{m'}}{2}} \; .
\end{equation}

By considering the opposite extreme twist $\kappa \to 0$ for the normalised twisted transfer matrix~\eqref{eq:Ttwistnormalized} we obtain another GT basis, giving eigenstates that diagonalise the D-operator, with $TQ$-relations~\eqref{eq:TQeffective} simplifying to
\begin{equation}
    D_{\mu,\vect{u} } (x) \, Q_{\mu,\vect{u}} (x) = D_\mu (x) \, Q_{\mu,\vect{u}} (x+1) \; .
\end{equation}
The solutions for the Bethe roots have a similar combinatorial structure as for $\kappa\to \infty$, with `partial strings' of consecutive zeroes that instead start from the right in the $\ell_m$-string.

\subsection{Norms and overlaps}
\label{subsec:descendantnorm}

The norms and overlaps of \textsc{yhw} states and their Yangian descendant states, i.e.\ the Bethe vectors, can be studied standard \textsc{aba} techniques.

\paragraph{Yangian highest-weight states}
To begin with, we prove that \textsc{yhw} states belonging to different motifs are orthogonal. Since the HS Hamiltonian is hermitian, eigenstates with different energies and different magnon numbers are automatically orthogonal. However, since the HS energy is so simple (integer in suitable units and additive on shell), different motifs may give rise to the same energy. Such `accidental degeneracies' do indeed occur~\cite{Finkel:2015bba, mathflow}. 
Instead, the eigenvalues \eqref{eq:Aeigenvaluehw} of the \textsc{yhw} states for the A--operator allow one to (uniquely) read off the motif. Thanks to \eqref{eq:ADconjugate} it follows that \textsc{yhw} states with different motifs $\mu$ and $\mu'$ are orthogonal as in \eqref{eq:pseudo_orthog}.

Let us now use the properties of Jack polynomials to give another proof of the orthogonality of the \textsc{yhw} states, and moreover determine their norms. 
The overlap between two \textsc{yhw} states with (possibly equal) motifs $\mu,\mu' \in\mathcal{M}_N$ is
\begin{equation}
    {_\mu}\braket{0}{0}_{\mu'} = \sum_{n_1 < \dots < n_M}^N \!\!\!\!\! \bigl| V(\omega^{n_1}, \dots\mspace{-1mu}, \omega^{n_M}) \bigr|^4 \, \mathsf{P}_{\nu(\mu)} (\omega^{-n_1}, \dots\mspace{-1mu}, \omega^{-n_M}) \,  \mathsf{P}_{\nu(\mu')} (\omega^{n_1} , \dots\mspace{-1mu}, \omega^{n_M} ) \; ,
\label{eq:Yhwnormstep1}
\end{equation}
where we recall that $\omega = \E^{2\rmi \pi/N}$ is the $N$th root of unity.
To convert the sum into a contour integral of the type that appears for the norm of Jack polynomials we use the identity
\begin{equation}
    \frac{1}{N} \sum_{k=1}^N (\omega^k)^n = \delta_{n,0} = \frac{1}{2\pi \rmi} \oint_{S^1} z^n \, \frac{\mathrm{d} z}{z} \; ,
\end{equation}
where the contour integral is along the unit circle $S^1$ in anticlockwise direction. This allows us to express the overlap \eqref{eq:Yhwnormstep1} as a multiple-contour integral,
\begin{equation}
\begin{aligned}
    {_\mu}\braket{0}{0}_{\mu'} = \frac{N^M }{(2\pi \rmi)^M\, M!} \oint_{S^1} \!\cdots \oint_{S^1} & \mathsf{P}_{\nu} \bigl(z_1^{-1} , \dots , z_M^{-1} \bigr) \, \mathsf{P}_{\nu'} \bigl(z_1 , \dots , z_M\bigr) \\
    & \times \prod_{m \neq m'}^M \biggl( 1 - \frac{z_m}{z_{m'}} \biggr)^{\!2} \prod_{m''=1}^M \! \frac{\mathrm{d} z_{m''}}{z_{m''}} \; ,
\end{aligned}
\end{equation}
where the Vandermonde part was rewritten using
\begin{equation}
    V(z_1 , \dots , z_M) \, V(z_1^{-1} , \dots , z_M^{-1} ) =  \prod_{m \neq m'}^M \biggl( 1 - \frac{z_m}{z_{m'}} \biggr) \; ,
\end{equation}
which is \eqref{eq:Phi0} with $\beta=2$. Its square should be understood as the measure for Jack polynomials with parameter $\alpha = 1/\beta = 1/2$, which are also known as spherical zonal polynomials. From the definition of the (Hall) inner product of these polynomials, using Eq.~(10.38) of Chapter~VI in \cite{macdonald}, the overlap takes the form
\begin{equation} \label{eq:hwnormderivation}
\begin{aligned}
   {_\mu}\braket{0}{0}_{\mu'} & = N^M \, \bigl\langle \mathsf{P}_{\nu(
   \mu)} , \mathsf{P}_{\nu(\mu')} \bigr\rangle'_{\!M} \\
    & = N^M \, \delta_{\nu(\mu) , \nu(\mu') } \! \prod_{m < m'}^M \!\! \frac{\Gamma (\xi_m - \xi_{m'} + \alpha^{-1} ) \, \Gamma (\xi_m - \xi_{m'} - \alpha^{-1} +1)}{\Gamma (\xi_m - \xi_{m'}) \, \Gamma (\xi_m - \xi_{m'} + 1)} \\
    & = N^M \, \delta_{\mu , \mu' } \! \prod_{m < m'}^M \frac{\mu_{M-m+1} - \mu_{M-m'+1} +1 }{\mu_{M-m+1} - \mu_{M-m'+1} -1} \\
    & = N^M \, \delta_{\mu , \mu' } \! \prod_{m < m'}^M \frac{\mu_{m'} - \mu_m + 1 }{\mu_{m'} - \mu_m -  1}\; ,
\end{aligned}
\end{equation}
where $\Gamma(\xi) = (\xi-1)!$ since $\alpha^{-1}=2$ and $\xi_m = \nu_m +(M-m)\,\alpha^{-1} = \mu_{M-m+1} -1$, cf.~\eqref{eq:nu_from_mu}, are integers. 
This once more proves the orthogonality \eqref{eq:pseudo_orthog} and establishes the norm formula~\eqref{eq:normHigh} for the \textsc{yhw} states. The latter can also be obtained by taking the freezing limit $\beta \to \infty$ in the more general results for the spin-CS system obtained by Takemura and Uglov~\cite{TU}.

\paragraph{Dual Bethe vectors.} 
For the Yangian descendant states, i.e.\ the Bethe vectors, we first need to introduce appropriate dual states. Like for the Heisenberg chain, the dual Bethe vectors are defined as
\begin{equation}
   {_\mu}\bra{\vect{u}} \coloneqq {_\mu}\bra{0} \prod_{k=1}^K \! \mathbf{C} (u_k) \propto \bigl( \, \ket{\vect{u}^\ast}_{\mu} \bigr)^\dagger \; .
\end{equation}
We emphasise that we consider $\kappa >0$, so that the eigenvectors and dual eigenvectors of $\mathbf{T}(x;\kappa)$ are orthogonal on shell, i.e.\ at solutions to the \textsc{bae}~\eqref{eq:effectiveBAE}. Moreover, the Bethe roots of on-shell Bethe vectors are real or come in complex conjugate pairs, i.e.\ the set of Bethe roots obeys $\vect{u}^* = \vect{u}$.

\paragraph{Off-shell orthogonality for different motifs.}
Thanks to the orthogonality of the \textsc{yhw} states, it is easy to see that Bethe vectors belonging to different multiplets are orthogonal. Indeed, for motifs $\mu,\mu' \in \mathcal{M}_N$ we find
\begin{equation}
\begin{split}
    {_{\mu'}}\braket{\vect{v}}{\vect{u}}_{\mu} & = {_{\mu'}}\bra{0} \prod_{k'=1}^{K} \! \mathbf{C}(v_{k'}) \, \prod_{k=1}^K \! \mathbf{B} (u_k) \, \ket{0}_{\mu} \\
    & = f\bigl( A_{\mu'} (\vect{v}) , D_{\mu'} (\vect{v}) , A_\mu (\vect{u}) , D_\mu (\vect{u}) \bigr) \ {_{\mu'}}\braket{0}{0}_{\mu} \ = 0 \; , \qquad \mu \neq \mu' \, ,
\end{split}
\end{equation}
where $f\bigl( A_{\mu'} (\vect{v}) , D_{\mu'} (\vect{v}) , A_\mu (\vect{u}) , D_\mu (\vect{u}) \bigr)$ is some polynomial in the eigenvalues of the A- and D-operators that arises due to the Yangian relations between the quantum operators (see App.~\ref{app:ABArelations}).

\paragraph{On-shell/off-shell overlaps.}
Suppose that the Bethe roots $\vect{u} = \{ u_1,\dots,u_K \}$ are on-shell, \textit{i.e.}\ satisfy the \textsc{bae} \eqref{eq:effectiveBAE} for a given motif $\mu \in \mathcal{M}_N$ and twist $\kappa > 0$.
The overlap of the resulting on-shell Bethe vector $\ket{\vect{u}}_\mu$ and an arbitrary off-shell Bethe (co)vector ${}_\mu \bra{\vect{v}}$ with $\vect{v} = \{v_1,\dots,v_{K'}\}$ and the same motif is given by the classic result of Slavnov~\cite{Slavnov:1989uvz} 
\begin{equation}
    \frac{{_\mu}\braket{\vect{v}}{\vect{u}}_{\mu}}{{_\mu}\braket{0}{0}_{\mu}}  = \delta_{K,K'} \Biggl( \, \prod_{k=1}^K A_\mu (v_k) \, D_\mu (u_k) \Biggr) \, S_{\vect{v}, \vect{u}} \; ,
\label{eq:Slavnovoverlap}
\end{equation}
where the Slavnov determinant is defined as
\begin{equation} \label{eq:Slavnov}
    S_{\vect{v} , \vect{u}} = \frac{\det\limits_{1\leqslant k,k' \leqslant K} \, \Omega ( u_k, v_{k'} )}{\det\limits_{1\leqslant k,k' \leqslant K} \displaystyle \frac{1}{u_k - v_{k'} + 1}} \; ,
\end{equation}
with
\begin{equation}
    \Omega (u,v) \coloneqq t(u-v) - t(v-u) \times \kappa^{-2} \, \frac{D_\mu (v)}{A_\mu (v)} \, \frac{Q_{\mu,\vect{u}} (v+1)}{Q_{\mu,\vect{u}} (v-1)} \; , \quad t(x) \coloneqq \frac{1}{x} - \frac{1}{x+1} \, ,
\end{equation}
and $Q_{\mu,\vect{u}} (x)$ the Baxter $Q$-polynomial~\eqref{eq:Q}.

\paragraph{Norms of on-shell Bethe vectors.}
The norm of Yangian descendant states is similarly given by the Gaudin formula~\cite{Gaudin_1981, Korepin:1982gg, Gaudin_2014}
\begin{equation}
    \frac{{_\mu}\braket{\vect{u}}{\vect{u}}_{\mu}}{{_\mu}\braket{0}{0}_{\mu} } 
    = \biggl( \, \prod_{k=1}^K A_\mu (u_k) \, D_\mu (u_k) \biggr) \, G_{\vect{u}, \vect{u}} \; ,
\end{equation}
where the Gaudin determinant is
\begin{equation} \label{eq:Gaudin}
    G_{\vect{u}, \vect{u}} \coloneqq \frac{\det\limits_{1\leqslant k,k' \leqslant K} \displaystyle \frac{\partial^2 \, Y_{\vect{u}, \vect{u}}}{\partial u_k \, \partial u_{k'}}}{\det\limits_{1\leqslant k,k' \leqslant K} \displaystyle  \frac{1}{u_k-u_{k'} -1}} \; .
\end{equation}
Here, the numerator features the Yang--Yang function 
\begin{equation}
\begin{aligned}
    Y_{\vect{u} , \vect{u}} = {} & \sum_{k=1}^K \sum_{n=1}^N \pigl( \bigl(u_k - A_\mu^{(n)}\bigr) \log\bigl(u_k - A_\mu^{(n)}\bigr) - \bigl(u_k - D_\mu^{(n)}\bigr) \log\bigl(u_k - D_\mu^{(n)}\bigr) \pigr) \\ 
    & - \sum_{k<k'}^K \pigl( (u_k - u_{k'} +1) \log (u_k - u_{k'} +1) - (u_k - u_{k'} -1) \log (u_k - u_{k'} -1) \pigr) \\
    & + \kappa^2 \sum_{k=1}^K u_k \; ,
\end{aligned}
\end{equation}
where the first line contains the zeroes of the polynomials $A_\mu(x)$ and $D_\mu(x)$ in \eqref{eq:Aeigenvaluehw}--\eqref{eq:Deigenvaluehw}, 
\begin{equation}
    A_\mu(x) = \prod_{n=1}^N \bigl(x- A_\mu^{(n)}\bigr) \, , \quad D_\mu(x) = \prod_{n=1}^N \bigl(x- D_\mu^{(n)}\bigr) \, .
\end{equation}
We recall that, by construction, the critical points of the Yang--Yang function are the logarithmic \textsc{bae}; in other words, the \textsc{bae} \eqref{eq:BAE_v2} correspond to equations 
\begin{equation}
    \exp \Biggl( \frac{\partial}{\partial u_k} Y_{\vect{u} , \vect{u}} \Biggr) = 1 \, , \qquad k=1,\dots,K \, .
\end{equation}

\section{Example: \texorpdfstring{$N=6$}{N=6}}
\label{sec:example}

\begin{figure}
    \centering
    \includegraphics[width=\linewidth]{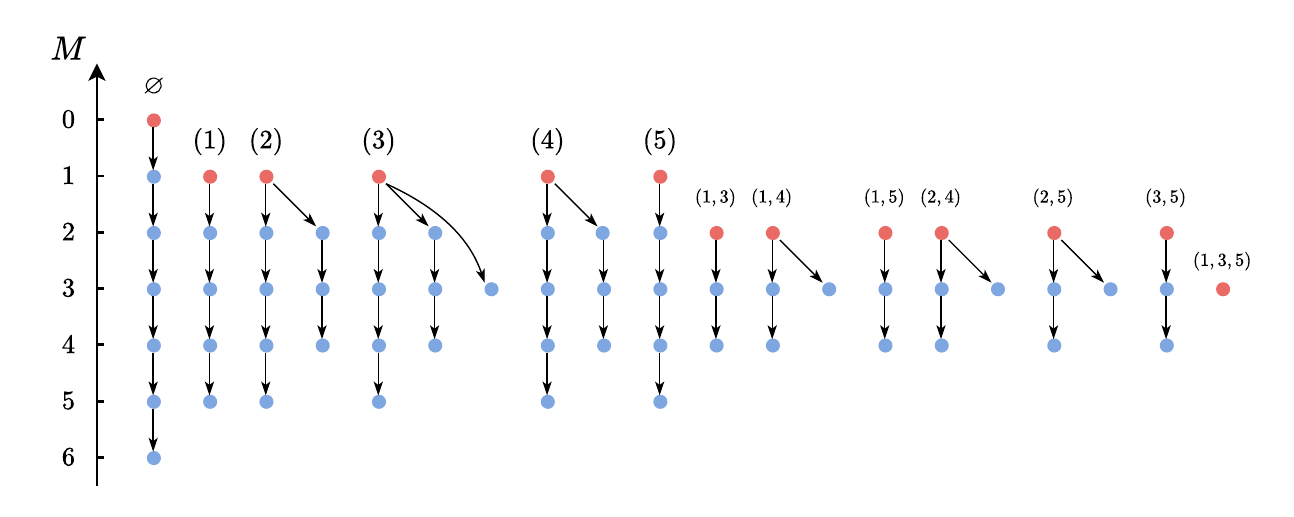}
    \caption{Schematic picture of the structure of the Hilbert space of the HS chain with $N=6$, with Yangian multiplets labelled by their motif (top) and total momentum (bottom), cf.~\cite[Fig.\,1]{LPS}. The red dots denote the \textsc{yhw} states, labelled by motifs, while the blue dots represent the descendant states. The vertically aligned states belong to the same $\mathfrak{sl}_2$ multiplets. The transfer matrix only preserves this structure in the periodic case (twist $\kappa=1$).}
    \label{fig:full_spec_N=6}
\end{figure}

To illustrate the above we work out the example $N=6$ in detail. In this case, there are 13 motifs 
\begin{equation}
    \mathcal{M}_6 = \{ \varnothing , (1) , (2) , (3) , (4) , (5) , (1,3) , (1,4) , (1,5) , (2,4) , (2,5) , (3,5) , (1,3,5) \} \; ,
\end{equation}
which label the eigenspaces of the HS chain, see Table~\ref{tb:N=6}. In the following, we will explain their structure as Yangian multiplets, i.e.\ the Yangian descendants of each \textsc{yhw} state $\ket{0}_\mu$. 
See also Fig.~\ref{fig:full_spec_N=6}.

\begin{table}[h]
    \centering
    \begin{tabular}{cccc|cc|c}
        $M$ & $\mu$ & $E$ & $p$ & $\bar{\nu}$ & $P_{\bar{\nu}}(\vect{z}) $ & spin \\ \hline
        $0$ & $\varnothing$ & $0$ & $0$ & $0$ & $1$ & $\textbf{3}$ \\
        $1$ & $(1)$ & $5/2$ & $\frac{\pi}{3}$ & $(0)$ & $1$ & $\hphantom{\textbf{0}\otimes } \textbf{0}\otimes \textbf{2} = \textbf{2} \hphantom{\oplus \textbf{1}\oplus \textbf{0}}$ \\[.2ex]
        & $(2)$ & $4$ & $\frac{2\pi}{3}$ & $(1)$ & $z_1$ & $\tfrac{\textbf{1}}{\textbf{2}}\otimes \textbf{0}\otimes \tfrac{\textbf{3}}{\textbf{2}} = \textbf{2}\oplus \textbf{1} \ \ \hphantom{\oplus \textbf{0}}$ \\
        & $(3)$ & $9/2$ & $\pi$ & $(2)$ & $z_1^2$ & $\textbf{1}\otimes \textbf{0}\otimes \textbf{1} = \textbf{2}\oplus \textbf{1}\oplus \textbf{0}$ \\
        & $(4)$ & $4$ & $\mathllap{-}\frac{2\pi}{3}$ & $(3)$ & $z_1^3$ & $\tfrac{\textbf{3}}{\textbf{2}} \otimes \textbf{0}\otimes \tfrac{\textbf{1}}{\textbf{2}} = \textbf{2}\oplus \textbf{1} \ \ \hphantom{\oplus \textbf{0}}$ \\[.2ex]
        & $(5)$ & $5/2$ & $\mathllap{-}\frac{\pi}{3}$ & $(4)$ & $z_1^4$ & $\textbf{2}\otimes \textbf{0} \hphantom{\otimes \textbf{0}} \ = \textbf{2} \hphantom{\oplus \textbf{1}\oplus \textbf{0}} \ $ \\[.2ex]
        $2$ & $(1,3)$ & $7$ & $\mathllap{-}\frac{2\pi}{3}$ & $(0,0)$ & $1$ & $\textbf{0}\otimes \textbf{0}\otimes \textbf{1} = \textbf{1}$ \\[.2ex]
         & $(1,4)$ & $13/2$ & $\mathllap{-}\frac{\pi}{3}$ & $(1,0)$ & $z_1 + z_2$ & $\textbf{0}\otimes \tfrac{\textbf{1}}{\textbf{2}} \otimes \textbf{0}\otimes \tfrac{\textbf{1}}{\textbf{2}} = \textbf{1} \oplus \textbf{0}$ \\[.2ex]
         & $(1,5)$ & $5$ & $0$ & $(2,0)$ & $z_1^2 + z_2^2 + \tfrac43 \, z_1\,z_2$ & $\textbf{0}\otimes \textbf{1} \otimes \textbf{0} = \textbf{1}$ \\
         & $(2,4)$ & $8$ & $0$ & $(1,1)$ & $z_1\,z_2$ & $\tfrac{\textbf{1}}{\textbf{2}} \otimes \textbf{0}\otimes \textbf{0}\otimes \tfrac{\textbf{1}}{\textbf{2}} = \textbf{1} \oplus \textbf{0}$ \\ 
        & $(2,5)$ & $13/2$ & $\frac{\pi}{3}$ & $(2,1)$ & $ z_1\,z_2 \, (z_1 + z_2)$ & $\tfrac{\textbf{1}}{\textbf{2}} \otimes \textbf{0}\otimes \tfrac{\textbf{1}}{\textbf{2}} \otimes \textbf{0} = \textbf{1} \oplus \textbf{0}$ \\[.2ex]
        & $(3,5)$ & $7$ & $\frac{2\pi}{3}$ & $(2,2)$ & $(z_1\,z_2)^2$ & $\textbf{1}\otimes \textbf{0} \otimes \textbf{0} = \textbf{1}$ \\
        $3$ & $(1,3,5)$ & $19/2$ & $\pi$ & $(0,0)$ & $1$ & $\textbf{0}\otimes \textbf{0} \otimes \textbf{0} = \textbf{0}$ \\
    \end{tabular}
    \caption{The motifs, HS energy, momentum, Jack polynomials in the \textsc{yhw} states, and $\mathfrak{sl}_2$-structure (where `$\textbf{s}$' means spin $s$) for all Yangian multiplets at $N=6$. We have $\bar{\nu}(\mu) = \nu(\mu) + 1$ compared to \eqref{eq:Y_hw_wf}.}
    \label{tb:N=6}
\end{table}

\subsection{Yangian highest-weight states}

When the zeroes of the Drinfeld polynomial consist of ($0$-strings and) a single $\ell$-string, the \textsc{yhw} state has only ordinary $\mathfrak{sl}_2$-descendants, which form a spin-$\frac{\ell}{2}$ multiplet, and there are no further affine descendants. The formula \eqref{eq:Drinfeldpoly} shows that this is the case for the motifs that are concentrated at one or both sides, i.e.\ are of the form $(2\,n-1)_{n=1}^a \cup (N-2n+1)_{n=1}^{b}$ for certain $a,b \in \mathbb{Z}_{\geqslant0}$. For $N=6$ these are
\begin{equation} \label{eq:N=6_motifs_onlysl2}
    N = 6 : \qquad \begin{array}{lcr}
    & {\varnothing} \, , & \\
    (1) \, , & & (5) \, , \\
    (1,3)\, , & \quad (1,5) \, , \quad & (3,5) \, , \\
    & (1,3,5) \, , &
    \end{array}
\end{equation}
see also Table~\ref{tb:N=6}. In particular, the antiferromagnetic ground state $\ket{0}_{(1,3,5)}$ is a singlet without any descendants. As always, the $\mathfrak{sl}_2$-descendants are obtained by acting with $\mathbf{S}^-$ on $\ket{0}_{\mu}$; in particular they are clearly independent of the twist $\kappa$. While the solutions to the TQ relation~\eqref{eq:TQeffective} do depend on $\kappa$, for the motifs in \eqref{eq:N=6_motifs_onlysl2} we have
\begin{equation}
    \ket{ \{u_1,\dots,u_K\} }_\mu \propto ( \mathbf{S}^-)^{\mspace{-2mu}K} \, \ket{0}_\mu
\end{equation}
when $\{u_1,\dots,u_K\}$ solves the TQ relation \eqref{eq:TQeffective} for a given twist $\kappa$ and motif $\mu$ from \eqref{eq:N=6_motifs_onlysl2}.

When the zeroes of a Drinfeld polynomial consist of more than one nontrivial $\ell$-string (i.e.\ with $\ell >0$), the corresponding \textsc{yhw} state has more than just $\mathfrak{sl}_2$-descendants. We call descendant states that are not mere $\mathfrak{sl}_2$-descendants \emph{affine descendants}; they require `affine' generators of the Yangian to be generated. In this case, the $\mathfrak{sl}_2$-descendants are typically not eigenvectors of the twisted transfer matrix, since the latter does not commute with $\mathbf{S}^-$ for $\kappa\neq 1$.

The remaining motifs, i.e.
\begin{equation}
    N = 6 : \qquad 
    \begin{gathered} 
    (2) \, ,\quad (3)\, ,\quad (4)\, ,\\ 
    (1,4) \, ,\quad (2,5) \, ,
    \end{gathered}
\end{equation}
label \textsc{yhw} states admitting affine descendants. See again Table~\ref{tb:N=6} and Fig.~\ref{fig:full_spec_N=6}.

\subsection{Descendants for various twists}

Considering the motif $\mu=(3)$ as an example, we describe its Yangian-descendant structure in turn for the GT limit, the periodic case and generic $\kappa$. 

The \textsc{yhw} state $\ket{0}_{(3)}$ is an eigenstate of the A- and D-operators, with eigenvalues \eqref{eq:Aeigenvaluehw}--\eqref{eq:Deigenvaluehw} given by
\begin{subequations}
    \begin{align}
    A_{(3)} (x) = \biggl(x+\frac{7}{2} \biggr) \biggl(x+\frac{5}{2} \biggr) \biggl(x+\frac{3}{2} \biggr) \biggl(x-\frac{1}{2} \biggr)^{\!2} \biggl(x-\frac{3}{2} \biggr) \; ,   \label{eq:eigenvalueA(3)} \\
    D_{(3)} (x) = \biggl(x+\frac{5}{2} \biggr) \biggl(x+\frac{3}{2} \biggr)^{\!2} \biggl(x-\frac{1}{2} \biggr) \biggl(x-\frac{3}{2} \biggr) \biggl(x-\frac{5}{2} \biggr) \; .  \label{eq:eigenvalueD(3)} 
    \end{align}
\end{subequations}
The corresponding Drinfeld polynomial is
\begin{equation} \label{eq:Dri_mu=3}
    P_{(3)} (x) = \biggl(x+ \frac{5}{2} \biggr) \biggl(x+ \frac{3}{2} \biggr) \biggl(x - \frac{3}{2} \biggr) \biggl(x - \frac{5}{2} \biggr) \; ,
\end{equation}
with zeroes $\{-5/2,-3/2\} \cup \{ 3/2,5/2\}$ forming two $2$-strings.

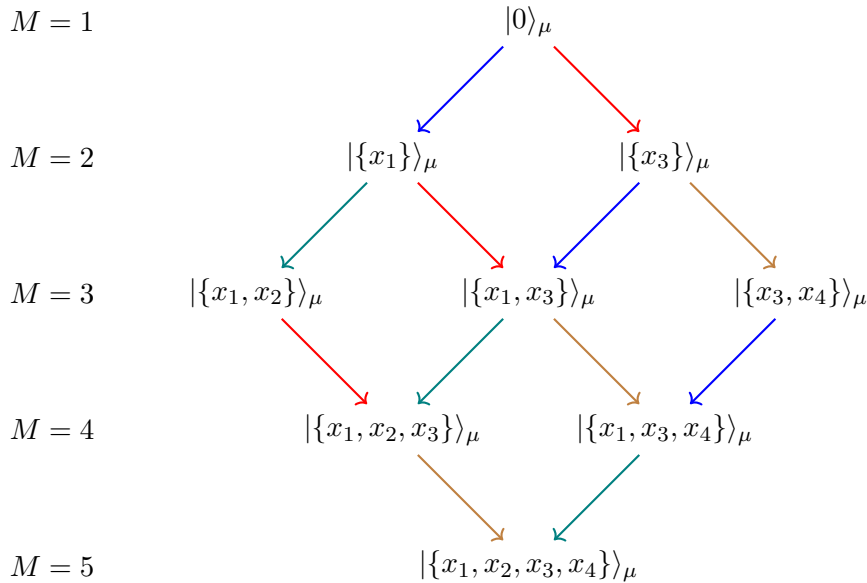
\begin{figure}[h]
    \centering
    \begin{tikzpicture}[scale=.9]
    \node (M1) at (-3,8) {$M=1$};
    \node (M2) at (-3,6) {$M=2$};
    \node (M3) at (-3,4) {$M=3$};
    \node (M4) at (-3,2) {$M=4$};
    \node (M5) at (-3,0) {$M=5$};
    \node (hw) at (4,8) {$|0 \rangle_{\mu}$};
    \node (des11) at (2,6) {$|\{x_1\} \rangle_{\mu}$};
    \node (des12) at (6,6) {$|\{x_3\} \rangle_{\mu}$};
    \node (des21) at (0,4) {$|\{x_1 , x_2\} \rangle_{\mu}$};
    \node (des22) at (4,4) {$|\{x_1,x_3\} \rangle_{\mu}$};
    \node (des23) at (8,4) {$|\{x_3,x_4\}\rangle_{\mu}$};
    \node (des31) at (2,2) {$|\{x_1 , x_2, x_3\}\rangle_{\mu}$};
    \node (des32) at (6,2) {$ |\{x_1 , x_3, x_4 \}\rangle_{\mu}$};
    \node (des4) at (4,0) {$|\{x_1,x_2,x_3,x_4\}\rangle_{\mu}$};
    \draw[color=blue, thick, ->] (hw) -- (des11);
    \draw[color=red, thick, ->] (hw) --  (des12);
    \draw[color=teal, thick, ->] (des11) --  (des21);
    \draw[color=red, thick, ->] (des11) -- (des22);
    \draw[color=blue, thick, ->] (des12) -- (des22);
    \draw[color=brown, thick, ->] (des12) -- (des23);
    \draw[color=red, thick, ->] (des21) --  (des31);
    \draw[color=teal, thick, ->] (des22) --  (des31);
    \draw[color=brown, thick, ->] (des22) --  (des32);
    \draw[color=blue, thick, ->] (des23) --  (des32);
    \draw[color=brown, thick, ->] (des31) --  (des4);
    \draw[color=teal, thick, ->] (des32) --  (des4);
    \end{tikzpicture}
    \caption{Hasse diagram showing the descendant structure for the Yangian multiplet for the motif $\mu = (3)$ at $N=6$ in the Gelfand--Tsetlin limit $\kappa\to\infty$. Different colours represent the action of B-operators with different arguments, with $x_1 = -\frac{7}{2}$ (\textcolor{blue}{blue}), $x_2 = -\frac{5}{2}$ (\textcolor{teal}{teal}), $x_3 = \frac{1}{2}$ (\textcolor{red}{red}) and $x_4 = \frac{3}{2}$ (\textcolor{brown}{brown}).}
    \label{fg:hasse_GT}
\end{figure}

\paragraph{Descendants in the GT limit.} 
While taking the limit $\kappa\to\infty$, the Yangian descendants become `GT descendants', which are all eigenstates of the A-operator. From the combinatorics of the GT basis described in Sec.~\ref{sec:GT}, the Bethe roots are determined by partial strings (starting from the left) of roots of the Drinfeld polynomial \eqref{eq:Dri_mu=3} shifted by $-1$. Starting with $K=1$ Bethe root, there are thus two GT descendants,
\begin{equation}
\begin{aligned}
    & \ket{\{ -\tfrac{7}{2} \}}_{(3)} \; \colon & A_{(3) , \{ -\frac{7}{2} \} } (x) & = \biggl(x+\frac{5}{2} \biggr)^{\!2} \biggl(x+\frac{3}{2} \biggr) \biggl(x-\frac{1}{2} \biggr)^2 \biggl(x-\frac{3}{2} \biggr) \; , \\
    & \ket{\{ \tfrac{1}{2} \}}_{(3)} \; \colon & A_{(3) , \{ \frac{1}{2} \} } (x) & = \biggl(x+\frac{7}{2} \biggr) \biggl(x+\frac{5}{2} \biggr) \biggl(x+\frac{3}{2} \biggr) \biggl(x-\frac{1}{2} \biggr) \biggl(x-\frac{3}{2} \biggr)^{\!2} \; , 
\end{aligned}
\end{equation}
where we also recorded the eigenvalues of the A-operator. At $K=2$ Bethe roots there are three GT descendants,
\begin{align}
    & \ket{ \{ -\tfrac{7}{2}, -\tfrac{5}{2} \} }_{(3) } \; \colon & A_{(3) , \{ -\frac{7}{2}, -\frac{5}{2} \} } (x) & = \biggl(x+\frac{5}{2} \biggr) \biggl(x+\frac{3}{2} \biggr)^{\!2} \biggl(x-\frac{1}{2} \biggr)^{\!2} \biggl(x-\frac{3}{2} \biggr) \; , \nonumber \\
    & \ket{ \{ -\tfrac{7}{2} , \tfrac{1}{2} \} }_{(3)} \; \colon & A_{(3) , \{ -\frac{7}{2}, \frac{1}{2} \} } (x) & = \biggl(x+\frac{5}{2} \biggr)^{\!2} \biggl(x+\frac{3}{2} \biggr) \biggl(x-\frac{1}{2} \biggr) \biggl(x-\frac{3}{2} \biggr)^{\!2} \; , \\
    & \ket{ \{ \tfrac{1}{2} , \tfrac{3}{2} \} }_{(3)} \; \colon & A_{(3) , \{ \frac{1}{2}, \frac{3}{2} \} } (x) & = \biggl(x+\frac{7}{2} \biggr) \biggl(x+\frac{5}{2} \biggr) \biggl(x+\frac{3}{2} \biggr) \biggl(x-\frac{1}{2} \biggr) \biggl(x-\frac{3}{2} \biggr) \biggl(x-\frac{5}{2} \biggr) \; . \nonumber
\end{align}
For $K=3$ there are again two GT descendants
\begin{align}
    & \ket{ \{ -\tfrac{7}{2}, -\tfrac{5}{2} , \tfrac{1}{2} \} }_{(3) } \; \colon & A_{(3) , \{ -\frac{7}{2}, -\frac{5}{2} , \frac{1}{2} \} } (x) & = \biggl(x+\frac{5}{2} \biggr) \biggl(x+\frac{3}{2} \biggr)^{\!2} \biggl(x - \frac{1}{2} \biggr) \biggl(x-\frac{3}{2} \biggr)^{\!2}  \; , \\ 
    & \ket{ \{ -\tfrac{7}{2}, \tfrac{1}{2} , \tfrac{3}{2} \} }_{(3)} \; \colon & A_{(3) , \{ -\frac{7}{2} , \frac{1}{2} , \frac{3}{2} \} } (x) & = \biggl(x+\frac{5}{2} \biggr)^{\!2} \biggl(x+\frac{3}{2} \biggr) \biggl(x-\frac{1}{2} \biggr) \biggl(x-\frac{3}{2} \biggr) \biggl(x-\frac{5}{2} \biggr) \; . \nonumber
\end{align}
Finally, there is one GT descendant with $K=4$, which is proportional to the counterpart of the \textsc{yhw} state beyond the equator,
\begin{equation}
    \ket{ \{ -\tfrac{7}{2}, -\tfrac{5}{2} ,\tfrac{1}{2} , \tfrac{3}{2} \} }_{(3)} \propto \prod_{j=1}^6 \! \sigma^x_j \; \ket{ 0 }_{(3)} \; : \quad 
     A_{(3) , \{ -\frac{7}{2}, -\frac{5}{2} , \frac{1}{2} , \frac{3}{2} \} } (x) 
     = D_{(3)} (x) \; .
\end{equation}     
This is the Yangian lowest-weight eigenstate, i.e.\ an eigenstate of both the A- and D-operator that is annihilated by the B-operator. The resulting structure of the Yangian multiplet can be organised in a Hasse diagram as shown in Fig.~\ref{fg:hasse_GT}, demonstrating the Gelfand--Tsetlin pattern~\cite{molev2007yangians}.

In general, the eigenvalues of the A-operator on the GT descendants have a simple relation to that for the \textsc{yhw} states,
\begin{equation}
    A_{\mu , \{ u_1, \dots, u_n \} } (x) = A_{\mu} (x) \prod_{m=1}^n \frac{x-u_m-1}{x-u_m} \; .
\end{equation}

\paragraph{Descendants with $\mathfrak{sl}_2$ symmetry.}
In the periodic limit $\kappa \to1$, the transfer matrix has additional $\mathfrak{sl}_2$ symmetry. From the Drinfeld polynomial \eqref{eq:Dri_mu=3} we read off the Clebsch--Gordon decomposition of the Yangian multiplet, viewed as a tensor product of $\mathfrak{sl}_2$-multiplets, as shown in Table~\ref{tb:N=6}. The two spin-$1$ representations correspond to the two $2$-strings in \eqref{eq:Dri_mu=3}, while the spin-$0$ representation corresponds to the contribution of the $0$-string. See also Fig.~\ref{fig:full_spec_N=6}. Therefore, there are three $\mathfrak{sl}_2$-multiplets: one containing the \textsc{yhw} state (spin $2$) and one each whose $\mathfrak{sl}_2$ highest-weight state has $K=1$ (spin $1$) or $K=2$ (spin $0$) additional Bethe roots.

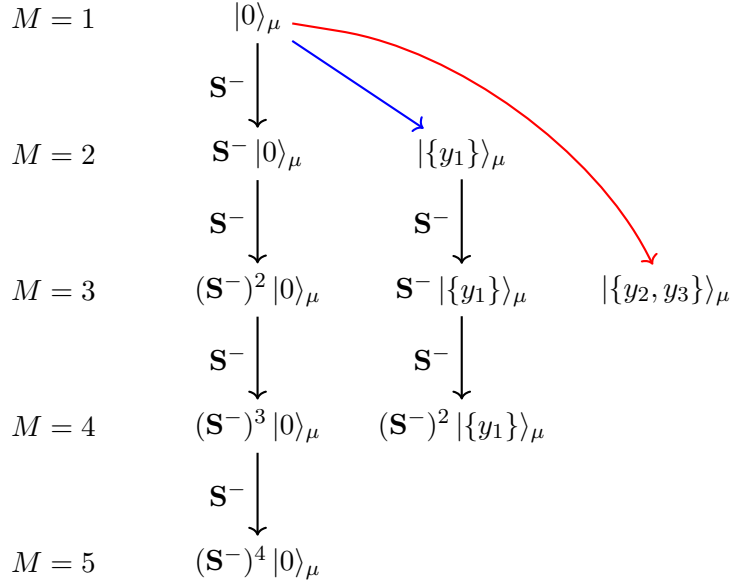
\begin{figure}[h]
    \centering
\begin{tikzpicture}[scale=.9]
    \node (M1) at (-3,8) {$M=1$};
    \node (M2) at (-3,6) {$M=2$};
    \node (M3) at (-3,4) {$M=3$};
    \node (M4) at (-3,2) {$M=4$};
    \node (M5) at (-3,0) {$M=5$};
    \node (hw) at (0,8) {$\ket{0}_{\mu}$};
    \node (hw1) at (0,6) {$\mathbf{S}^- \, \ket{0}_{\mu}$};
    \node (hw2) at (0,4) {$(\mathbf{S}^-)^2 \, \ket{0}_{\mu}$};
    \node (hw3) at (0,2) {$(\mathbf{S}^-)^3 \, \ket{0}_{\mu}$};
    \node (hw4) at (0,0) {$(\mathbf{S}^-)^4 \, \ket{0}_{\mu}$};
    \node (des1) at (3,6) {$|\{y_1\}\rangle_{\mu}$};
    \node (des11) at (3,4) {$\mathbf{S}^- \, |\{y_1\}\rangle_{\mu}$};
    \node (des12) at (3,2) {$(\mathbf{S}^-)^2 \, |\{y_1\}\rangle_{\mu}$};
    \node (des2) at (6,4) {$|\{y_2,y_3\}\rangle_{\mu}$};
    \draw[color=black, thick, ->] (hw) -- node[midway, left] {$\mathbf{S}^-$} (hw1);
    \draw[color=black, thick, ->] (hw1) -- node[midway, left] {$\mathbf{S}^-$} (hw2);
    \draw[color=black, thick, ->] (hw2) -- node[midway, left] {$\mathbf{S}^-$} (hw3);
    \draw[color=black, thick, ->] (hw3) -- node[midway, left] {$\mathbf{S}^-$} (hw4);
    \draw[color=blue, thick, ->] (hw) -- (des1);
    \draw[color=black, thick, ->] (des1) -- node[midway, left] {$\mathbf{S}^-$} (des11);
    \draw[color=black, thick, ->] (des11) -- node[midway, left] {$\mathbf{S}^-$} (des12);
    \draw[color=red, thick, rounded corners=80pt, ->] (hw) -- (4.3,7.3) -- (des2);
\end{tikzpicture}
    \caption{Counterpart for Fig.~\ref{fg:hasse_GT} with $\kappa=1$. Again $N=6$ and $\mu = (3)$. Here $y_1 = -\frac{1}{2}$ (\textcolor{blue}{blue}) and the complex conjugate pair $y_2 = y_3^\ast = -\frac{1}{2} + \rmi$ (\textcolor{red}{red}). Like the \textsc{yhw} state $\ket{0}_\mu$ (at the top of the quintet), the affine descendants $|\{y_1\} \rangle_{\mu}$ (at the top of the triplet) and $|\{ y_2, y_3\} \rangle_{\mu}$ (a singlet) have $\mathfrak{sl}_2$ highest weight.}
    \label{fg:hasse_ka=1}
\end{figure}

Solving the TQ relation~\eqref{eq:TQeffective} for the spin-$1$ $\mathfrak{sl}_2$ highest-weight state we have
\begin{equation}
    |\{ -\tfrac{1}{2} \} \rangle_{(3) } \; ,
\end{equation}
and for the spin-$0$ $\mathfrak{sl}_2$ highest-weight state we have
\begin{equation}
    |\{ -\tfrac{1}{2} - \mathrm{i} , -\tfrac{1}{2} + \mathrm{i} \} \rangle_{(3)} \; .
\end{equation}
Taking into account all $\mathfrak{sl}_2$ descendants we arrive at one spin-$2$, one spin-$1$, and one spin-$0$ $\mathfrak{sl}_2$ multiplet,
\begin{equation}
    \begin{gathered}
    \ket{0}_{(3) } \, , \qquad\mathbf{S}^- \, \ket{0}_{(3) } \, , \qquad ( \mathbf{S}^-)^2 \, \ket{0}_{(3) } \, ,\qquad ( \mathbf{S}^-)^3 \, \ket{0}_{(3) }\, ,\qquad ( \mathbf{S}^- )^4\, \ket{0}_{(3) } \, ; \\
    |\{ -\tfrac{1}{2} \} \rangle_{(3)} \, , \qquad \mathbf{S}^- \,  |\{ -\tfrac{1}{2} \} \rangle_{(3) } \, , \quad ( \mathbf{S}^-)^2 \, |\{ -\tfrac{1}{2} \} \rangle_{(3)} \, ; \quad \\
    |\{ -\tfrac{1}{2} - \mathrm{i} , -\tfrac{1}{2} + \mathrm{i} \} \rangle_{(3) } \; . \quad
    \end{gathered}
\end{equation}
This structure of the Yangian multiplet is analogous to that of the usual Heisenberg \textsc{xxx} spin chain. A Hasse diagram for this Yangian-descendant tower with $\kappa=1$ is shown in Fig.~\ref{fg:hasse_ka=1}. The descendant structure for the whole Hilbert space is as shown in Fig.~\ref{fig:full_spec_N=6}.

\paragraph{Descendants at generic twist.} 
In this subsection, we present the numerical solutions of the Yangian descendants of the \textsc{yhw} state $|0\rangle_{(3)}$ with a generic twist, which we take to be $\kappa = \frac{1}{\sqrt{2}}$ for definiteness.

For Yangian descendants with $K=1$ additional magnon, we solve the effective $TQ$-relation~\eqref{eq:TQeffective} to find two physical solutions,
\begin{equation}
    \ket{ \{ \tfrac{5+4\sqrt{3}}{2} \} }_{(3) } \; ,  \quad \ket{ \{ \tfrac{5-4\sqrt{3}}{2} \} }_{(3) } \; .
\end{equation}
In the absence of closed-form Bethe roots, we present the numerical solutions to 
six digits in the following. Since all non-real Bethe roots come in complex conjugate pairs~($\kappa>0$), we use the shorthand $u\pm v\,\rmi$ to mean $u+ v\,\rmi,u- v\,\rmi$.
For Bethe vectors with $K=2$ additional magnons we find three physical solutions,
\begin{equation}
    \ket{ \{ 4.35245 \pm 2.23356 \, \rmi \} }_{(3) } \; , \quad 
    \ket{ \{ -1.55338 \pm 0.519379 \, \rmi \} }_{(3)} \; , \quad
    \ket{ \{ -0.698802, 4.10066\} }_{(3)} \; .
\label{eq:generickappaM3}
\end{equation}
For $K=3$ we find two physical solutions,
\begin{equation}
    \ket{ \{ 3.81572 , 2.07419 \pm 2.98985 \, \rmi \} }_{(3) } \; , \quad
    \ket{ \{3.34600 , -1.15505 \pm 0.908272 \, \rmi \} }_{(3) } \; .
\end{equation}
Finally, for Yangian descendants with $K=4$ additional magnons, there is only one physical solution, that gives the Yangian lowest-weight state
\begin{equation}
\begin{split}
    \ket{ \{  -0.61811 \pm 2.84188 \, \rmi , 2.61811 \pm 0.812557 \, \rmi \} }_{( 3 ) } \propto \prod_{n=1}^6 \! \sigma^x_n \; \ket{ 0 }_{(3) } \; .
\end{split}
\end{equation}

This illustrates that for a generic twist $\kappa$, the Yangian descendants associated with the twisted transfer matrix $\mathbf{T}(x;\kappa)$ also provide the correct number of states in the corresponding Yangian multiplet. This time, the Bethe roots of the different Yangian descendants are unrelated, unlike for the special cases $\kappa\to\infty$ and $\kappa=1$. Just like for (e.g.\ twisted \textsc{xxx}, or \textsc{xxz}) Heisenberg chains without $\mathfrak{su}_2$ symmetry, there is no extra structure in the Yangian descendant tower.

\paragraph{Comparison.} 
For a fixed magnon number $M$, the descendant states for different $\kappa$ span the same Hilbert space. Different choices of $\kappa$ imply different choices of the basis states. For instance, the descendant states with $M=3$ in \eqref{eq:generickappaM3} span the same Hilbert space as
$|\{x_1, x_2\} \rangle_\mu , |\{x_1, x_3\} \rangle_\mu , |\{x_3, x_4\} \rangle_\mu$
in Fig.~\ref{fg:hasse_GT} span in the limit $\kappa \to \infty$, and as
$(\mathbf{S}^-)^2 |0\rangle_\mu , \mathbf{S}^- |\{y_1\}\rangle_\mu , |\{y_1, y_2\}\rangle_\mu$
in Fig.~\ref{fg:hasse_ka=1} span when $\kappa = 1$. 
These form eigenbases for different extra Heisenberg-style symmetries \eqref{eq:Heis_style} or \eqref{eq:Heis_style_isotr}, depending on the twist, but are all valid eigenbases for the HS hamiltonian~\eqref{eq:H_HS}.

\section{Conclusion and outlook}
\label{sec:conclusion}

This work provides a detailed and systematic construction of Yangian descendant states for the Haldane--Shastry (HS) chain capitalising on \cite{FLLS}. Since each HS eigenspace can be separately viewed as an `effective' inhomogeneous Heisenberg \textsc{xxx} chain, we use an `internal Bethe ansatz' following \cite{FLLS} by formulating these descendant states as eigenstates of the twisted transfer matrix $\mathbf{T}(x; \kappa)$ with twist $\kappa>0$ to ensure hermiticity. Starting from the explicitly known \textsc{yhw} state for a given motif, we derive the \textsc{bae} and TQ-relations for the Bethe roots of its Yangian descendants at given $\kappa$. With the help of the algebraic Bethe ansatz (\textsc{aba}), the eigenvalues and eigenvectors of the transfer matrix are obtained. The quantum operators require one to go through the spin-CS system and freezing, cf.\ the end of Sec.~\ref{sec:CSbasis}, meaning that the construction of the explicit Bethe vectors requires some effort. Nevertheless, the \textsc{aba} description enables us to compute their norms and overlaps, yielding compact determinant formulae in terms of the Bethe roots. 
In the extreme cases $\kappa\to\infty$ and $\kappa\to 0$~(the Gelfand--Tsetlin limits), the associated $TQ$-relation simplifies considerably and becomes explicitly solvable. 

It will be interesting to connect our spin-chain approach to the \textsc{cft} techniques of \cite{Cirac:2010rka, Herwerth:2015pga} through the known realisation of the Yangian symmetry in the \textsc{cft}~\cite{HHTBP,Bernard:1994wg, Bouwknegt:1994bk, Bouwknegt:1994sj, Schoutens:1994au}.

Our results establish a foundation for computing a wide range of physical observables. An immediate and important goal is the analytical study of quantum quench dynamics. In integrable spin chains with nearest-neighbour interaction, such dynamics for integrable initial states~\cite{Caux_2013, Brockmann_2014, Caux:2016esd, Piroli:2017sei, Pozsgay:2018dzs}, including the Néel state and integrable matrix-product states, have been central to the understanding of non-equilibrium behaviour~\cite{Caux_2013, Ilievski_2015}. Extending similar analytical approaches to long-range interacting systems is highly desirable, particularly given the experimental relevance of such models in platforms such as ion traps. In this context, exact overlap formulas between eigenstates (typically expressed as Bethe states) and integrable boundary states serve as essential inputs. Systematic studies of integrable boundary states as well as their overlaps with energy eigenstates are still lacking for long-range spin chains. The present work constitutes a crucial step toward this direction. With information of the descendant states, we also expect our results to provide a useful basis for advancing research towards finite-temperature correlations.

Another interesting direction is the investigation of physical properties at finite temperature. While the thermodynamics of the HS chain has been explored from several perspectives, first via the free-spinon-gas picture~\cite{Haldane_1991_spinon_gas} and later through more conventional thermodynamic approaches~\cite{Ha1993}, the relationship between these descriptions has recently been revisited in~\cite{Bulchandani_2024}. Moreover, hydrodynamic methods have been applied to examine transport properties in this model~\cite{Bulchandani_2024}. These studies collectively highlight the analytical tractability of the HS chain, for which many physical quantities admit closed-form expressions. A key open problem in this area is the microscopic derivation of the Boltzmann equation; addressing it will require a deeper understanding of the Hilbert-space structure and descendant states studied here.

\section*{Acknowledgement}

We are indebted to Jean-Sébastien Caux for collaboration in the early stages of this work and for numerous discussions throughout the course of this work. We thank Didina Serban for helpful discussions and useful references. We are grateful to Hong-Hao Tu for informing us about the results of \cite{Cirac:2010rka, Herwerth:2015pga} and discussions. Y.M.\ thanks Jiakang Bao, Henry Liu and Masahito Yamazaki for useful discussions on the Yangian. Y.M.\ is grateful for the hospitality of Rudolf Peierls Centre for Theoretical Physics and St.~John's College at University of Oxford. The work of Y.M.\ was supported by the World Premier International Research Center Initiative (WPI), MEXT, Japan and the UTokyo Global Activity Support Program for Young Researchers. The work of Y.J.\ is supported by National Natural Science Foundation of China through Grant No.12575073. Y.M.\ used ChatGPT 5.5 to help read \cite{TU}; we acknowledge the use of ChatGPT 5.5 to assist with proofreading the manuscript. The authors of this paper are ordered alphabetically.

\appendix

\section{Yangian relations of quantum operators}
\label{app:ABArelations}

In this appendix, for easy reference we collect some of the well-known non-trivial algebraic relations of quantum operators $\mathbf{A}(x)$, $\mathbf{B}(x)$, $\mathbf{C}(x)$ and $\mathbf{D}(x)$ contained in the RTT relation \eqref{eq:RMM}. These are useful when deriving the \textsc{bae} (or $TQ$-relation) as well as the norm and overlap formulae. The relations read 
\begin{equation}
    \left[ \mathbf{A}(x) , \mathbf{A}(y) \right] = \left[ \mathbf{B}(x) , \mathbf{B}(y) \right] = \left[ \mathbf{C}(x) , \mathbf{C}(y) \right] = \left[ \mathbf{D}(x) , \mathbf{D}(y) \right] = 0 \, ,
\end{equation}
\begin{align}
    & \left[ \mathbf{A}(x) , \mathbf{D}(y) \right] = \frac{1}{x-y} \, \bigl( \mathbf{C}(y) \, \mathbf{B}(x) - \mathbf{C}(x) \, \mathbf{B}(y) \bigr) \; , \\
    & \left[ \mathbf{D}(x) , \mathbf{A}(y) \right] = \frac{1}{x-y} \, \bigl( \mathbf{B}(y) \, \mathbf{C}(x) - \mathbf{B}(x) \, \mathbf{C}(y) \bigr) \; , \\
    & \left[ \mathbf{C}(x) , \mathbf{B}(y) \right] = \frac{1}{x-y} \, \bigl( \mathbf{A}(y) \, \mathbf{D}(x) - \mathbf{A}(x) \, \mathbf{D}(y) \bigr) \; , \\
    & \left[ \mathbf{B}(x) , \mathbf{C}(y) \right] = \frac{1}{x-y} \, \bigl( \mathbf{D}(y) \, \mathbf{A}(x) - \mathbf{D}(x) \, \mathbf{A}(y) \bigr) \; , 
\end{align}
\begin{align}
    & \mathbf{A}(y) \, \mathbf{B} (x) = \frac{x-y}{x-y+1} \, \mathbf{B} (x) \, \mathbf{A} (y) + \frac{1}{y-x} \, \mathbf{B} (y) \, \mathbf{A} (x) \; , \\
    & \mathbf{B}(y) \, \mathbf{A} (x) = \frac{x-y}{x-y+1} \, \mathbf{A} (x) \, \mathbf{B} (y) + \frac{1}{y-x} \, \mathbf{A} (y) \, \mathbf{B} (x) \; , \\
    & \mathbf{D}(y) \, \mathbf{C} (x) = \frac{x-y}{x-y+1} \, \mathbf{C} (x) \, \mathbf{D} (y) + \frac{1}{y-x} \, \mathbf{C} (y) \, \mathbf{D} (x) \; , \\
    & \mathbf{C}(y) \, \mathbf{D} (x) = \frac{x-y}{x-y+1} \, \mathbf{D} (x) \, \mathbf{C} (y) + \frac{1}{y-x} \, \mathbf{D} (y) \, \mathbf{C} (x) \; , 
\end{align}
\begin{align}
    & \mathbf{A}(x) \, \mathbf{C} (y) = \frac{x-y}{x-y+1} \, \mathbf{C} (y) \, \mathbf{A} (x) + \frac{1}{y-x} \, \mathbf{C} (x) \, \mathbf{A} (y) \; , \\
    & \mathbf{C}(x) \, \mathbf{A} (y) = \frac{x-y}{x-y+1} \, \mathbf{A} (y) \, \mathbf{C} (x) + \frac{1}{y-x} \, \mathbf{A} (x) \, \mathbf{C} (y) \; , \\
    & \mathbf{B}(x) \, \mathbf{D} (y) = \frac{x-y}{x-y+1} \, \mathbf{D} (y) \, \mathbf{B} (x) + \frac{1}{y-x} \, \mathbf{D} (x) \, \mathbf{B} (y) \; , \\
    & \mathbf{D}(x) \, \mathbf{B} (y) = \frac{x-y}{x-y+1} \, \mathbf{B} (y) \, \mathbf{D} (x) + \frac{1}{y-x} \, \mathbf{B} (x) \, \mathbf{D} (y) \; . 
\end{align}

\section{Inhomogeneous Heisenberg \textsc{xxx} chain}
\label{app:inhomogeneousXXX}

In this appendix, we consider an inhomogeneous Heisenberg \textsc{xxx} chain with linearly increasing inhomogeneities. We will show that it contains a similar Yangian descendant structure as the HS chain.

We define the inhomogeneous monodromy and transfer matrices as 
\begin{equation}
\begin{aligned}
    & \mathbf{M}_{a}^{\rm inh} (x) = \mathbf{R}_{a1}\Bigl(x-\xi_1 -\tfrac{1}{2}\Bigr) \cdots \mathbf{R}_{aN} \Bigl(x-\xi_N - \tfrac{1}{2}\Bigr) = 
    \begin{pmatrix} 
    \mathbf{A}^{\rm inh}(x) & \mathbf{B}^{\rm inh}(x) \\ 
    \mathbf{C}^{\rm inh}(x) & \mathbf{D}^{\rm inh}(x) \end{pmatrix}_{\!a} \; , \\
    & \mathbf{T}^{\rm inh} (x ; \kappa) = \kappa \,\mathbf{A}^{\rm inh} (x) + \kappa^{-1} \, \mathbf{D}^{\rm inh} (x) \; .
\end{aligned}
\end{equation}
Following \textsection4.3.2 of \cite{FLLS}, we choose the following (maximally non-generic) inhomogeneities
\begin{equation}
    \xi = \{ -\tfrac{N-1}{2} , -\tfrac{N-3}{2} , \cdots , \tfrac{N-1}{2} \} \; ,
\end{equation}
such that all neighbouring inhomogeneities differ by $\xi_n - \xi_{n-1} = 1$, meaning that we encounter (antisymmetric) fusion at each pair of sites, see e.g.\ \textsection2.2.4 in \cite{FLLS}. In particular, one important observation is that given a sequence of integers $\mu = \{ \mu_1, \dots , \mu_M\} $ obeying the motif conditions~\eqref{eq:motifs}, the state
\begin{equation} \label{eq:inh_mu}
    \ket{\mu}^{\rm inh} \coloneqq \prod_{m=1}^M \! \mathbf{B} \Bigl(\mu_m -\tfrac{N+1}{2}\Bigr) \, \ket{\uparrow\cdots\uparrow} \propto 
    \prod_{m=1}^M \! \bigl( \sigma_{\mu_m + 1}^- - \sigma_{\mu_m}^- \bigr) \, \ket{\uparrow\cdots\uparrow} \,
\end{equation}
has singlets $\ket{\uparrow\downarrow}-\ket{\downarrow\uparrow}$ at pairs of neighbouring sites $(\mu_m,\mu_m+1)$ indicated by the motif. Moreover, the state $\ket{\mu}^\mathrm{inh}$ has \textsc{yhw},
\begin{equation}
    \begin{pmatrix}
        \mathbf{A}^\mathrm{inh}(x) & \mathbf{B}^\mathrm{inh}(x) \\ \mathbf{C}^\mathrm{inh}(x) & \mathbf{D}^\mathrm{inh}(x)
    \end{pmatrix}_{\!a} \ket{\mu}^\mathrm{inh} = 
    \begin{pmatrix}
        A_\mu(x) \, \ket{\mu}^\mathrm{inh} & \ast \\ 0 & D_\mu(x) \, \ket{\mu}^\mathrm{inh} 
    \end{pmatrix}_{\!a} \, ,
\end{equation}
with the same eigenvalues \eqref{eq:Aeigenvaluehw}--\eqref{eq:Deigenvaluehw} as for the \textsc{yhw} state $\ket{0}_\mu$ in the HS chain. We refer the readers to \cite{FLLS} for a more detailed explanation.

\subsection{The Q-system}
\label{app:Qsystem}

Solving the \textsc{bae} or $TQ$-relation~\eqref{eq:TQeffective} for the inhomogeneous Heisenberg \textsc{xxx} chain leads to both physical and unphysical solutions. To get a (complete) set of Bethe roots of inhomogeneous \textsc{xxx} chain that precisely correspond to the physical states, one can solve the rational Q-system. We briefly review the Q-system in this appendix. For more details and examples we refer to \cite{Marboe:2016yyn, Jiang:2017phk, Bohm:2022ata}.

\begin{figure}[h]
    \centering
    \begin{tikzpicture}[arrowlabel/.style={gray},gridline/.style={black, line width=1.1pt},boundarynode/.style={circle,draw=black,line width=0.5pt,minimum size=2mm,inner sep=0pt},scale=.8]
    
    \draw[gridline] (0,1) rectangle (4,2);
    \foreach \x in {1,2,3}
        \draw[gridline] (\x,1) -- (\x,2);
    
    \draw[gridline] (0,0) rectangle (6,1);
    \foreach \x in {1,2,3,4,5}
        \draw[gridline] (\x,0) -- (\x,1);
    
    \foreach \x in {0,...,6} {
        \node[boundarynode, fill=gray!25] at (\x,0) {};
        \node[boundarynode, fill=gray!25] at (\x,1) {};
        };
    \foreach \x in {0,...,4} \node[boundarynode, fill=gray!25] at (\x,2) {};
    
    \foreach \x in {0,...,4}
        \node at (\x,2) [above,yshift=.1cm] {$1$};
    
    \node at (0,1) [left] {$Q_{1,0}$};
    \node at (0,0) [below left] {$Q_{0,0}\!\!$};
    \node at (1,0) [below] {$Q_{0,1}$};
    \node at (2.5,0) [below] {$\vphantom{Q_{0,1}}\cdots$};
    \node at (4,0) [below] {$Q_{0,M}$};
    \node at (5,0) [below] {$\ \vphantom{Q_{0,1}}\cdots$};
    \node at (6,0) [below right] {$\!Q_{0,L-M}$};
    \node at (6,1) [above right] {$1$};
    \node at (5,1) [above] {$\ \vphantom{Q_{0,1}}\cdots$};
    \end{tikzpicture}
    \caption{Graphical representation of the Q-system attached to a Young diagram $(L-M,M)$.}
    \label{fig:Qsystem}
\end{figure}
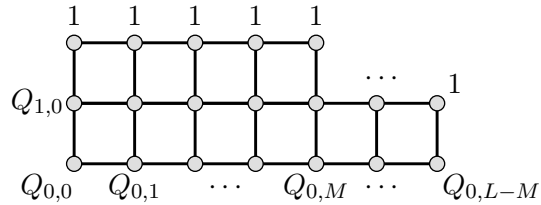

For the (inhomogeneous) \textsc{bae} with length $L$ and $M$ magnons, the Q-system is associated to a two-row Young diagram $(L-M,M)$, illustrated in Fig.~\ref{fig:Qsystem}. We use the French notation for the Young diagram, i.e.\ the longer row is at the bottom. At each node of the Young diagram we define a function of the spectral parameter $u$, denoted by $Q_{a,n}(u)$ where $(a,n)$ is the lattice coordinate of the node on the Young diagram, cf.~Fig.~\ref{fig:Qsystem}. The Q-functions are not independent: the Q-functions at the four corners of each box are related by the QQ-relation
\begin{equation} \label{eq:rationalQsys}
    \begin{aligned}
    Q_{0,n} \, Q_{1,n-1} & = \kappa \, Q_{0,n-1}^+ \, Q_{1,n}^- - \kappa^{-1} \, Q_{0,n-1}^- \, Q_{1,n}^+ \;, \\
     Q_{1,n} & = Q_{1,n-1}^+ - Q_{1,n-1}^- \; , 
    \end{aligned}
    \qquad n \geqslant 1 \;,
\end{equation}
where $\kappa$ is the twist in the transfer matrix and we have fixed $Q_{2,n}=1$ as part of the boundary conditions. We use the notation $f^{\pm}(x) \coloneqq f(x \pm \frac{1}{2})$ to denote the shift of argument of a function. The Q-functions at the left boundary are also completely or partly fixed. In particular, we define 
\begin{equation}
    Q_{0,0} (x) = \prod_{n=1}^N (x- \xi_n) = \prod_{n=1}^N \Bigl( x + \tfrac{N+1}{2}-n \Bigr) \; , 
\end{equation}
which is completely fixed. 

The Baxter's Q-function $Q_{1,0}$ is fixed to take the following form
\begin{equation}
    \begin{aligned}
    Q_{1,0} (x) = Q_{\mu,\vect{u}} \Bigl(x - \tfrac{1}{2} \Bigr) & = \prod_{m=1}^M \! \Bigl( x- \tfrac{N+1-2\,\mu_m}{2} - \tfrac{1}{2} \Bigr) \prod_{k=1}^{K} \! \Bigl( x- u_k - \tfrac{1}{2} \Bigr) \\
    & = \prod_{m=1}^M \! \Bigl( x- \tfrac{N+1-2\,\mu_m}{2} - \tfrac{1}{2} \Bigr) \left(x^{K} + \sum_{k=0}^{K-1}c_k \, x^k \right) \; ,
    \end{aligned}
\end{equation}
for Yangian pseudovacuum $|\mu \rangle^{\rm inh}$. 
Note that $Q_{1,0}(x)$ is only partly fixed because we still need to find the Bethe roots $\{u_k\}_{k=1}^{K}$. We further impose the analytic condition that all the Q-functions on the Young diagrams are polynomials in $x$. This leads to a set of equations for the coefficients $c_0,\dots,c_{K-1}$ called zero-remainder conditions. Solving these equations for the coefficients $c_k$ gives the polynomial $Q_{1,0}(x)$ and thus its zeroes, i.e.\ the Bethe roots. 

By solving the rational Q-system \eqref{eq:rationalQsys}, we obtain precisely all physical solutions to the inhomogeneous Heisenberg \textsc{xxx} chain. 

The solutions to the Q-system are in one-to-one correspondence with the HS eigenstates. The Yangian descendants for the HS chain can be obtained by acting with $\prod_{k=1}^K \mathbf{B}(u_k)$ on the corresponding \textsc{yhw} state $\ket{0}_\mu$. For a fixed value of the twist~$\kappa$, the $TQ$-relations \eqref{eq:TQeffective} of the HS chain are equivalent to the QQ-relations \eqref{eq:rationalQsys} with $n=1$. 

However, an important difference between the inhomogeneous \textsc{xxx} chain and HS chain is the form of the \textsc{yhw} states. For the inhomogeneous \textsc{xxx} chain, these consists of $\uparrow$s combined with singlets realising the motif in real space, and can be obtained via acting with the B-operator as in \eqref{eq:inh_mu}. In contrast, the \textsc{yhw} states of the HS chain are given in terms of Jack polynomials, and cannot be obtained simply by acting with B-operators on the ferromagnetic vacuum $\ket{\uparrow \cdots \uparrow}$, and the motifs rather manifest themselves in momentum (or spectral) space, see Sec.~\ref{sec:motifs}. Nevertheless, the eigenvalues of the A- and D-operators on the \textsc{yhw} states for the two spin chains coincide for a given motif $\mu$.

\section{Reduced Yangian generators for a given motif}
\label{app:reducedYangian}

In this appendix we define a rescaled monodromy matrix whose B-operator does not vanish at special Bethe roots in the Gelfand--Tsetlin limit (extreme twist) discussed in Sec.~\ref{sec:GT}. In the language of \cite{FLLS}, this is the monodromy matrix of the `effective' spin chain describing an eigenspace of the HS chain.

Fix a motif $\mu$. We define the polynomial $g_{\mu} (x)$ as the greatest common divisor of the eigenvalues \eqref{eq:Aeigenvaluehw}--\eqref{eq:Deigenvaluehw} of the A- and D-operators on the \textsc{yhw} state $\ket{0}_\mu$,
\begin{equation} \label{eq:gcd mu}
	g_{\mu} (x) \coloneqq \mathrm{gcd} \bigl( A_\mu (x) , D_\mu (x) \bigr) \; . 
\end{equation}
In view of the expression \eqref{eq:Drinfeldpoly} of the Drinfeld polynomial, $g_{\mu} (x)$ consists of two factors,
\begin{equation}
	g_{\mu} (x) = S_\mu (x) \; \mathrm{gcd}\bigl( P_\mu (x) , P_\mu (x+1) \bigr) \; ,
\end{equation}
where
\begin{equation}
	S_\mu (x) \coloneqq \prod_{n \in \mu} \biggl( x+ \frac{N+1-2(n-1)}{2} \biggr) \biggl( x+ \frac{N+1-2(n+1)}{2} \biggr)
\end{equation}
is the contribution of the antisymmetric fusion to singlets as in \eqref{eq:inh_mu}, while the greatest common divisor of $P_\mu (x)$ and $P_\mu (x+1)$ accounts for the contribution of symmetric fusion into spin-$\frac{\ell_m}{2}$ for $\ell_m\geq 2$, $m = 0,\dots, M$. See also the discussion of degeneracies in Sec.~\ref{subsec:yangian}.

For this motif $\mu$, we define the \emph{reduced} monodromy matrix
\begin{equation}
	\mathbf{M}^\mathrm{red}_{\mu,\,a} (x) \coloneqq \frac{1}{g_\mu(x)} \, \mathbf{M}_a (x) = 
    \begin{pmatrix} \mathbf{A}^\mathrm{red}_\mu (x) & \mathbf{B}^\mathrm{red} _\mu (x) \\ \mathbf{C}^\mathrm{red}_\mu (x) & \mathbf{D}^\mathrm{red}_\mu (x) 
    \end{pmatrix}_{\!a} \; ,
\end{equation}
which also acts on the Yangian irrep labelled by the motif $\mu$. Concretely, on the \textsc{yhw} state $\ket{0}_\mu$, we have 
\begin{equation}
	\mathbf{A}^\mathrm{red}_\mu (x) \, \ket{0}_{\mu} = \frac{A_\mu(x)}{g_\mu (x)} \, \ket{0}_{\mu} \; , \quad \mathbf{D}^\mathrm{red}_\mu (x) \, \ket{0}_{\mu} = \frac{D_\mu(x)}{g_\mu (x)} \, \ket{0}_{\mu} \; ,
\end{equation}
where, due to the definition \eqref{eq:gcd mu} of $g_\mu$, both eigenvalues are polynomials in $x$ of degree $N - \mathrm{deg}(g_\mu)$. Crucially, the resulting Drinfeld polynomial $P_\mu(x)$ remains the same, see \eqref{eq:A_over_D_vs_Dri}. Therefore the same Drinfeld polynomial $P_\mu(x)$ yields a Yangian structure that is isomorphic. This reduced monodromy matrix is `minimal' for the motif $\mu$ in the sense that its eigenvalues have lowest degree while still corresponding to $P_\mu(x)$. It can be viewed as defining an inhomogeneous Heisenberg \textsc{xxx} chain with at most $M+1$ sites carrying spins $\ell_0/2,\ell_1/2,\dots,\ell_M/2$ ignoring any $\ell_m = 0$; this is the `effective spin chain' in the terminology of \cite{FLLS}.

Therefore, the Yangian descendants within the multiplet with motif $\mu$ may conveniently be defined using the reduced B-operator,
\begin{equation}
	\ket{\vect{u}}_\mu^{\rm red} = \prod_{u \in \vect{u}} \!\mathbf{B}^\mathrm{red}(u) \, \ket{0}_\mu \; ,
\end{equation}
whose norm and overlap formulae are obtained by renormalising the results from Sec.~\ref{subsec:descendantnorm}.
The advantage of the reduced B-operator is that for any Bethe root $u_0$ such that $g_\mu(u_0) = 0$, as occurs in \eqref{eq:GT_from_ABA}, the renormalised descendant state
\begin{equation}
	\mathbf{B}^\mathrm{red} (u_0) \, \ket{\vect{u}}_\mu^{\rm red}
\end{equation}
is nonzero, whereas $\mathbf{B}(u_0) \, \ket{\vect{u}}_\mu =0$. This is the reason why we use the reduced B-operators to construct the \textsc{gt} descendants in \eqref{eq:GT_from_ABA}.

\bibliography{BiBTeX_File.bib}

\end{document}